# Black Hole Paradoxes

**Mario Rabinowitz**
Armor Research
715 Lakemead Way
Redwood City, CA 94062-3922
Mario715@earthlink.net

**Abstract**

The Black Hole enigma has produced many paradoxes since its inception by John Michell in 1783. As each paradox was resolved and our understanding about black holes matured, new paradoxes arose. A consensus regarding the resolution of some conundrums such as the Naked Singularity Paradox and the Black Hole Lost Information Paradox (LIP) has still not been achieved. Black hole complementarity as related to the LIP and the LIP itself are challenged by gravitational tunneling radiation. Where possible, the paradoxes will be presented in historical context presenting the interplay of competing perspectives such as those of Bekenstein, Belinski, Chandrasekar, Finkelstein, Hawking, Maldacena, Page, Penrose, Preskill, Susskind, 't Hooft, Veneziano, Wald, Winterberg, Yilmaz, and others. The simplest possible equation $G\rho\hbar/90$ is obtained for Hawking radiation. The average kinetic energy of emitted particles may have a feature in common with thermionic emission. A broad range of topics will be covered including: Why can or can't the formation of a black hole be observed? Can one observe a naked singularity like the one clothed by a black hole? What can come out of, or seem to come out of, a black hole? What happens to the information that falls into a black hole? Doesn't the resolution of the original black hole entropy paradox introduce an equally challenging puzzle? Related anomalies in the speed of light, the speed of gravity, the speed of inflation, and Mach's principle are considered. Black hole entropy violation of the Nernst heat theorem is virtual As we shall see, these paradoxes have served as an incentive for research to sharpen our thinking, and even to promote the development of new theoretical and experimental physics. A reasonable number of paradoxes are a good thing. But too many paradoxes requiring manifold patchwork fixes may augur basic flaws/inconsistencies that signal the need for a fundamental paradigm shift.

**1 Introduction**
**2 Einstein General Relativity and Black Hole Collapse**
  **2.1 The Schwarzschild metric**
  **2.2 Black hole collapse is unobservable**
  **2.3 Different solutions of the black hole observation paradox**
**3 Naked Singularity**

# 1  Introduction

Everything is connected in some way with everything else.  I have yet to find a scientific concept that came into being in isolation, independent of its predecessors.  The first recorded concept of a body whose gravitational field is so strong that not even light can leave it, can be traced to John Michell (1784).  Michell found the conditions for a body to have an escape velocity equal to the velocity of light, $c = 3 \times 10^{10}$ cm/sec = 186,000 mi/sec ≈ 1 ft/nanosec.  It is noteworthy that Michell derived exactly the same expression for the radius $R_H$ of a black hole of mass M, as obtained from general relativity. By conservation of energy in Newtonian physics,

$$\tfrac{1}{2} mc^2 - \frac{GMm}{R_H} = KE_\infty + PE_\infty = 0 \Rightarrow R_H = \frac{2GM}{c^2}, \qquad (1.1)$$

where G is Newton's universal gravitational constant. The kinetic energy term on the LHS is written non-relativistically here as it is in other conventional non-general relativistic derivations to avoid being a factor of 2 low for $R_H$. This simplified derivation is used to avoid undue complexity. It differs from general relativity in allowing emitted particles to be classically found at varying distances from a black hole. In general relativity, classically there are no emitted particles outside a black hole.

The velocity of light was determined a century earlier by the astronomer Roemer in 1675 from observations on the moons of Jupiter, and an application of Kepler's laws. That the velocity of light is finite was considered by Galileo in 1638. He and his assistant attempted to measure the speed of light by signaling with shuttered lanterns one mile apart. When the assistant saw light from Galileo's lantern, he unveiled his lantern, and Galileo tried to measure the time interval. He correctly concluded that the velocity was faster than they could measure. In hindsight, we know that a two-mile round trip for light would take only about 10 microseconds --much, much faster than they could measure.

Michell created the first black hole paradox, that the heavens can be filled with hot stars that cannot be seen because their light cannot reach us due to their high gravitational fields. Pierre Simon Laplace (1799) rediscovered this black hole paradox saying, "Therefore there exists, in the immensity of space, opaque bodies as considerable in magnitude, and perhaps equally as numerous as the stars." However, he retracted his work in light of Young's two-slit interference experiment and Huygens' wave theory of light since he had used a Newtonian particle (photon) approach in his derivation.

Subrahmanyan Chandrasekar (1931) proposed the first mechanism for the creation of black holes from stellar evolution -- causing a rift between him and his mentor Arthur Eddington, who didn't believe stars could collapse and become black holes. Chandrasekar showed that black holes were the final destiny of all stars with mass $\geq 1.4\, M_{sun}$. That is why all white dwarf stars have mass $< 1.4\, M_{sun}$. Chandrasekar confronted the antinomy that electron degeneracy pressure of the Pauli Exclusion Principle might prevent the otherwise irresistible gravitational collapse of a star. He showed that the gravitational field of a sufficiently massive star could subvert even the Pauli Exclusion Principle. It took decades for the astronomical community to accept his findings.

The early emphasis on large black holes is historical. Stephen Hawking (1971) was the first to propose that the high energies of the Big Bang created LBH (little black holes) in the primordial universe. Putting the creation of LBH in the era of the Big Bang resolves the disparity that they cannot be created by gravitational collapse in the present universe.

Albert Einstein (1915) published a short communication version of his Theory of General Relativity, one of the most profound theories conceived by the human mind. In two papers less than two months later, Karl Schwarzschild (1916) derived a spherically symmetric solution of Einstein's General Relativity for a black hole at rest, which he requested Einstein to submit for him. It is remarkable that not only did Schwarzschild do this so quickly, but that he did it while fatally ill with pemphigus (a debilitating disease with burning blisters on the skin and mucous membranes), while in active combat in WWI for the German Army on the Russian Front (Chandrasekar, 1983). Einstein presented Schwarzschild's solution on his behalf on Jan. 13, 1916 to the Prussian Academy. Unfortunately, Schwarzschild died June 19, 1916 **--**too soon to be aware of the importance of his work.

On an astronomical scale, black holes are the centers of attraction of galaxies, and generate the vast power emitted by quasars, the most luminous objects known in the universe -- or so most astronomers and astrophysicists believe. Quasars (quasi-stellar objects) were discovered in 1963. These celestial bodies are considered to have been formed 15 billion years ago at the beginning of the universe. Quasars are an extreme form of active galactic nuclei, powered by supermassive black holes of $10^6$ - $10^{10}$ solar masses, whose luminosity (power output) far exceeds the luminosity of their entire galaxy (Davies, 1992).

The above examples are only the tip of the iceberg in showing the importance of the concept of black holes. Let us look deeper and see how the black hole concept has raised anomalies in many other areas such as entropy, radiation, clothed or naked singularities, and information loss. This chapter focuses on the enigmas engendered by black holes and is not intended to present a comprehensive review of the many subjects covered -- that would be a book in itself. Only a small fraction of the seminal papers can be covered.

## 2  Einstein General Relativity and Black Hole Collapse

### 2.1  The Schwarzschild metric

In Einstein's General Relativity (EGR), motion is along geodesics. A geodesic (the shortest distance between two points in a given space) is a geometrical property independent of the physical characteristics of the moving body. Schwarzschild (1916) derived a spherically symmetric solution of EGR for an uncharged black hole of mass M with no angular momentum. The Schwarzschild metric is

$$ds^2 = -[1-\frac{2GM}{c^2r}]c^2dt^2 + [1-\frac{2GM}{c^2r}]^{-1}dr^2 + r^2[d\Theta^2 + \sin^2\Theta d\Phi^2], \quad (2.1)$$

where units are usually chosen so that G, and c can be set = 1. Since the metric coefficients are explicitly independent of time and there is no dragging of the inertial frame, the space-time is static for an observer outside the black hole and t is the proper time if the observer is at rest at infinity.

Friedwardt Winterberg (2002) presented a simple heuristic derivation of the Schwarzschild metric. Even though it is not rigorous, it does provide insight into eq. (2.1) in terms of Newtonian gravity (NG) and special relativity. It is presented here with some minor additions. By conservation of energy, a body of mass m acquires a velocity v when falling in the Newtonian gravitational field of mass M:

$$\tfrac{1}{2}mv^2 - \frac{GMm}{r} = 0 \Rightarrow v^2 = \frac{2GM}{r}. \quad (2.2)$$

The kinetic energy term on the LHS is written non-relativistically here and for eq. (1.1) as it is in other conventional non-general relativistic derivations to avoid being a factor of 2 low for $R_H$.

Combining eq. (2.2) with the length contraction of special relativity

$$dr = dr'\sqrt{1-\frac{v^2}{c^2}} = dr'\sqrt{1-\frac{2GM}{c^2r}}. \quad (2.3)$$

Combining eq. (2.2) with the time dilation of special relativity

$$dt = \frac{dt'}{\sqrt{1-\frac{v^2}{c^2}}} = \frac{dt'}{\sqrt{1-\frac{2GM}{c^2r}}}. \quad (2.4)$$

A distant observer far from M effectively at r ~ ∞ measures the unprimed variables such as dr and dt, whereas the primed variables such as dr' and dt' are measured in the rest frame of the body falling toward M.

The line element of special relativity is

$$ds^2 = dr'^2 - c^2 dt'^2 = dr^2 - c^2 dt^2. \tag{2.5}$$

Substituting eqs. (2.3) and (2.4) into the primed part of eqs. (2.5)

$$ds^2 = \frac{dr^2}{1 - \frac{2GM}{c^2 r}} - c^2 \left(1 - \frac{2GM}{c^2 r}\right) dt^2. \tag{2.6}$$

Special relativity is for Euclidean (flat) space. For spherical symmetry, with motion in the radial direction, eq. (2.6) becomes the Schwarzchild metric

$$ds^2 = \frac{dr^2}{1 - \frac{2GM}{c^2 r}} - c^2 \left(1 - \frac{2GM}{c^2 r}\right) dt^2 + r^2 \left[d\Theta^2 + \sin^2\Theta d\Phi^2\right], \tag{2.1}$$

where $\Theta$ is the polar angle, and $\Phi$ is the azimuthal angle. The singularities at $r = 2GM/c^2$ are virtual, and can be transformed away by a different choice of coordinate system. The singularity at $r = 0$ is real, and cannot be transformed away. In the weak field limit, eq. (2.1) reduces to the line element of special relativity given by eq. (2.5).

**2.2 Black hole collapse is unobservable**

The boundary radius of the black hole, called the Schwarzschild radius or the horizon radius is at

$$R_H = \frac{2GM}{c^2}. \tag{2.7}$$

At this radius the falling body reaches the velocity $v = c$. As can be seen from eq. (2.4), a distant observer sees time slowing down for the falling body as $v$ approaches $c$, until at $v = c$ time is frozen. Therefore, the time for the body to reach $R_H$ is infinite as seen by a distant observer. The same holds true for the collapse of a star due to its own gravitational field. No one will see it go inside $R_H$, or even reach $R_H$. This despite the fact that the star -- or any body falling into a black hole -- keeps going right through $R_H$. From the orthodox point of view, collapse of a body into a black hole is unobservable, and unstoppable once it starts.

The term horizon radius was coined by Wolfgang Rindler. Just as we cannot see below the horizon of the earth, a distant observer cannot see a star implode below $R_H$. Nevertheless, the solution of Schwarzschild's equation dictates that the star will inexorably continue to shrink down to r = 0, notwithstanding the fact that no one outside $R_H$ will be able to observe this. One can see this from different points of view. One perspective is that as v approaches c, the light from the body Doppler shifts to zero frequency, i.e. infinite wavelength making it impossible to see the body.

Another point of view is that the radiation from the falling body -- or the black hole itself for that matter -- undergoes a gravitational redshift. Particles that originate at or outside the horizon of an isolated black hole must lose energy in escaping the gravitational potential of the black hole. For example, in the case of photons

$$\nu_{obs} = \nu(r_{emission\ site})\left[1 - \frac{2GM/c^2}{r}\right]^{1/2} = \nu(r_{emiss\ site})\left[1 - \frac{R_H}{r}\right]^{1/2} \quad (2.8)$$

where $\nu_{obs}$ is the frequency detected by the observer at a very large distance from the black hole, and $\nu(r_{emission\ site})$ is the frequency at the radial distance r from the center of the black hole where the photon was created outside the black hole. Equation (2.8) does not depend on the gravitational potential between the emission site and observer, but only on the emission proximity to $R_H$. We see from eq. (2.8) that for r = $R_H$ and any finite $\nu(r_{emission\ site})$, then $\nu_{observer} = 0$ implying that any finite temperature at $R_H$ must red shift to zero as measured at large distances from a black hole.

## 2.3 Different solutions of the black hole observation paradox

Many prominent doubting physicists were reluctant to accept the concept of a black hole, in large measure because of the anomaly that the implosion of a star cannot be observed -- even in principle -- at and below $R_H$. Yet a particle inexorably falls into the center at r = 0 -- just as time marches on -- because in general relativity time and space exchange roles inside a black hole (which has the making of another paradox). One resolution, or at least reconciliation, of the observational black hole paradox was achieved only in the last few decades. David Finkelstein (1958) ascertained a new reference frame or framework for Schwarzschild space-time. Finkelstein's reference frame simultaneously covered all regions of space-time from the imploding star to distant reaches of space. This all-embracing reference frame is now called the Eddington-Finkelstein frame. This reference frame assuaged the concerns of the doubters, indicating that a star really does collapse below $R_H$, but only appears to stop at $R_H$ to distant observers.

Winterberg has taken a different approach which completely avoids the observation paradox . In his theory (Winterberg, 2002), as the energy of a particle approaches the Planck energy it starts to obey Newtonian physics (NP) rather than EGR. ['t Hooft also thinks there is a need for an underlying deterministic physics underpinning quantum mechanics at the Planck scale, cf. Sec 4.3.] When a star collapses, the particle energies readily approach the Planck energy. In EGR the red shift leads to an infinite time for a distant observer to see the star collapse through the black hole horizon. There is a less severe red shift in NP, and in Winterberg's model a distant observer can follow the gravitational implosion of a star ~ the mass of our sun in ~ $10^{-4}$ sec through and inside $R_H$, instead of the infinite time given by EGR. In both his model and in EGR, the black hole starts to form at the center of the imploding star. For Winterberg, a black hole develops at $r_{star} = 0$ when the star has collapsed to

$$r_{collapse} = \tfrac{3}{2} R_H . \qquad (2.9)$$

In EGR (Tolman, 1958) black hole formation starts at the center of the star, when the radius of the star reaches

$$r_{collapse} = \tfrac{9}{8} R_H . \qquad (2.10)$$

## 3 Naked Singularity

### 3.1 Cosmic censorship conjecture

Roger Penrose (1969) coined the cosmic censorship conjecture:

> We are thus presented with what is perhaps the most fundamental unanswered question of general-relativistic collapse theory, namely: does there exist a "cosmic censor" who forbids the appearance to distant observers of naked singularities, clothing each one in an absolute event horizon?

Put in simple black hole language, this says that singularities caused by gravitational collapse must remain invisible to distant observers. Even light emitted by a star inside the event horizon, $R_H$, would not only be confined within $R_H$, but like matter its trajectory would shrink to a singularity.  These are regions of space-time where the laws of physics break down due to infinite curvature, infinite mass density, infinite pressure, infinite temperature, etc.

Such singularities are distinct from the those that may be caused momentarily by the disappearance of a black hole by Hawking radiation (Hawking, 1974, 1975).  The production of such singularities falls in the realm of quantum gravity, and as such is likely to be prohibited.  In any case, black hole radiation is a quantum mechanical process and cosmic censorship refers only to classical EGR.  So, as Hawking put it, if black hole radiation were to produce a naked singularity, cosmic censorship would be transcended rather than violated.

It is not too surprising that Penrose (1979) in his treatise on Singularities and Time-Asymmetry examined the connection between entropy and the "arrow of time."  What may be unexpected is his statement relating singularities and the "arrow of time."   He also presented a stronger version of the cosmic censorship question:

> ...the presence and apparent structure of spacetime singularities contain the key to the solution to one of the long-standing mysteries of physics:  the origin of the "arrow of time."  ... Apart from a possible initial singularity (such as the "big bang" singularity) no singularity is ever "visible" to any observer.

This is a much stronger version than the 1969 version because it is intended to restrict observation of a naked singularity to any observer in any space-time -- not just distant observers in asymptotically flat space-time.

There is a need for a precise mathematical formulation for the problem of possible cosmic censorship.  Wald (1984) points out that one must specify what conditions the matter fields must obey.  For example two natural conditions are that the stress-energy tensor must satisfy an energy condition; and the coupled EGR equations must allow for a well-defined initial value condition.  However, perfect fluids (continuous and incompressible) obey both of these conditions, yet can violate cosmic censorship because even in flat space-time their dynamical evolution can result in singularities such as those caused by the formation of shock waves. Wald goes on to pose precise mathematical formulations for both forms of the cosmic censorship conjecture.

For Wald:

> The issue of whether the cosmic censor conjecture is correct remains the key unresolved issue in the theory of gravitational collapse.  The physical relevance of black holes depends in large measure on the validity of this conjecture.

I would agree with him on the first point. It appears that Wald is saying in his second point that if naked singularities can be observed, black holes are physically irrelevant.

I think that if black holes exist, they are relevant independent of whether the cosmic censor conjecture is correct or not. Their relevance is only slightly diminished if black holes can't hide all singularities. Since gravitational redshift does not allow one to see inside black holes, one might think that for a similar reason one would be prevented from observing a singularity. Let us see why this may not be true.

**3.2 Eluding cosmic censorship**

In Newtonian space-time, the concept of a singularity is clear-cut, it is a place in space and time where a physical quantity becomes infinite. The concept of a singularity is not nearly as intuitive in EGR in part because time and place are obtained *a posteriori*, rather than being given *a priori*. A wide variety of singularities, both real and virtual, can occur in EGR, so it is difficult to give a plain meaning to the concept. For example, in a strict interpretation of EGR, the big bang singularity is not considered to be part of the space-time manifold. It occurs at neither a place or a time. Similarly, from a rigorous point of view, the singularity at r = 0 inside a black hole is not a place. However, for our purposes, we will take a heuristic approach using NG examples for intuitive illustration.

Winterberg's model (2002) would have no trouble with the observation of a naked singularity inside $R_H$, if one existed. This is not a violation of cosmic censorship because his approach is not standard EGR. Just as his model permits observation of the implosion of a star inside $R_H$, it would potentially permit observation of a naked singularity. However, Winterberg has taken care to avoid any kind of singularity in what he calls his "finitistic" approach since he only deals with non-transcendental numbers such as integers and fractions.

For him numbers like pi are only a representation of the ratio of the perimeter of an inscribed or circumscribed regular polygon to the radius of a circle. He thinks physics should be formulated in terms of finite difference operators instead of differential equations. In his theory all intervals in space and time can only be measured in integer multiples of the Planck length and the Planck time. Although this sounds like "loop quantum gravity," in other respects he parts company with it.

It may be possible to elude cosmic censorship even within the context of EGR. EGR predicts the formation of singularities in regions of space in which the very presence of singularities challenges its efficacy. Observationally EGR could be safe for two reasons:

1) It may take an infinite amount of time to observe the formation of a singularity.

2) Cosmic censorship is valid, and singularities cannot be observed in principle for any reason.

The challenge for finding a way to avoid cosmic censorship is great, and many have tried. Stuart Shapiro and Saul Teukolsky (1991 a, b) appear to have been the first to find a computer simulation of a naked singularity in the collapse of a cluster of particles in the form of a prolate spheroid, which they call a spindle singularity. However as they emphasize, to completely rule out the possibility of the formation of an event horizon that would veil the naked singularity, they would have to map space-time far into the future. However as soon as they encounter a singularity, their numerical simulation encumbers the computer to the point of early termination.

There is also another source of concern related to their numerical computation. In NG, conservation of angular momentum speeds up the cluster of particles as they fall in toward the rotational axis parallel to the major axis of the prolate spheroid. No matter how small the original angular velocity, this speed up prevents the particles from being crushed to an infinitesimally thin singularity along the rotational axis. Without angular momentum, Newtonian theory predicts the formation of a line singularity. All singularities in Newtonian gravity are naked. EGR is more complicated, and the jury is out on this question. However for reasons of computer capacity, Shapiro and Teukolsky's model lacked physical reality because it had no angular momentum.

Nevertheless, the physics behind their model is clear. An intuitive insight in Newtonian terms, is also qualitatively valid in EGR. Since matter is concentrated only along the major axis, the gravitational field approaches infinity as $1/r$, thus being weaker than the point singularity field attained by a spherically symmetric distribution of matter which approaches infinity as $1/r^2$. It is interesting to note that Khalatnikov and Lifshitz (1970) thought they had proved that a spherical distribution of matter with random perturbations could not, in EGR, implode into a singularity for reasons of angular momentum. But this was incorrect.

In NG, a spherical shell of matter with no angular momentum will mutually attract until the matter all reaches $r = 0$ at the same time, producing a singularity. However with the slightest angular momentum or asymmetry, no singularity can form. Newton argued that an infinite universe would not collapse to a singularity due to gravity, because an infinite universe would have no point that could serve as the center of attraction since every point has an infinite number of points on all sides of it. His view will be examined in Sec. 8.7.

Let us see why, even in the simple case of a line mass singularity in non-relativistic NG, the question of the radius where the escape velocity = c is not trivial. As we saw in the derivation of eq. (1.1) for the black hole radius of a spherical mass M, NG emulates EGR well. By Gauss's law in three-dimensional Euclidean space

$$\oint \vec{F} \cdot d\vec{A} = F(2\pi rL) = -4\pi GMm \Rightarrow F = \frac{-2Gm\lambda}{r}, \tag{3.1}$$

where F is the force on mass m due to a line mass M of linear mass density $\lambda = M/L$. Since the force is minus the potential energy gradient:

$$\vec{F} = -\vec{\nabla}V \Rightarrow V_r - V_a = 2Gm\lambda \ln(r/a). \tag{3.2}$$

We can now anticipate a potential problem here, since the 0 of potential energy is at r = a, rather than at infinity (as required by EGR). To make matters worse, the potential energy is infinite at infinity, rather than 0 as in the case of a spherical distribution of mass. By conservation of energy,

$$\tfrac{1}{2}mv_a^2 + V_a = \tfrac{1}{2}mv_r^2 + V_r \Rightarrow \tfrac{1}{2}m\left(v_a^2 - v_r^2\right) = 2Gm\lambda \ln(r/a). \tag{3.3}$$

In non-relativistic NG, we want a particle of mass m to escape to r = ∞ for an escape velocity $v_a$ = c, where it converts all of its initial kinetic energy $\tfrac{1}{2}mv_a^2 = \tfrac{1}{2}mc^2$ at a = $R_H$ to potential energy at infinity, with $v_r = 0$. But a finite initial kinetic energy cannot be converted to an infinite potential energy. If we try to get $R_H$ simply by substituting into eq. (3.3) this yields

$$R_H = r \exp\left| \frac{-\left(c^2 - v_r^2\right)}{4G\lambda} \right| \xrightarrow[r \to \infty]{} \infty, \tag{3.4}$$

which seems self-contradictory as it says that the entire space is a Newtonian black hole. With an infinite speed of light, this could be circumvented, yielding $R_H$ = 0. Such strange behavior does not come unexpectedly. Equation (3.3) tells us that the line mass singularity cannot be escaped arbitrarily far away, with any finite initial velocity at any finite initial radius. An infinitely long line was chosen as a simple illustration, but part of the problem is that such a line is an infinite source with peculiarities of its own.

Hüseyin Yilmaz eludes Einsteinian black holes in general relativity and hence the cosmic censorship question for a different reason than attempted by Khalatnikov and Lifshitz, or Winterberg. Yilmaz (1958, 1982) developed an interesting variation of EGR --we'll call it YGR) which avoids Einsteinian black holes altogether as described in Sec. 9.

<u>Local</u> energy density is not well defined in EGR, and may not be a meaningful concept. This is because the space-time metric defines both the background space-time structure and the dynamical aspects of the gravitational field, but these two parts cannot be separated in the metric. Energy must be attributed to the dynamical aspect of gravity such as gravity waves, but may not be expected in the background static space-time structure. Nevertheless, in EGR the <u>total</u> energy of an isolated system can be defined by the gravitational field at infinity, and the radiated flux of gravitational energy is well defined.

EGR ascribes positive energy to gravitational radiation in the far field. Just as electromagnetic waves can take on an existence independent of the accelerated charges that produced them, quadrupole gravitational radiation can exist in space-time independent of the accelerated matter that created it. So in EGR, gravitational radiation in space-time that is free of matter can in principle collapse to form black holes.

Low energy (long wavelength) gravitational waves have only a small attraction between them, passing freely through one another and then dispersing. For high energy (short wavelength) gravity waves, their mutual attraction causes them to converge. In EGR, convergence begets more convergence in a runaway effect that ultimately leads to the production of a singularity that most likely is clothed in a black hole horizon. Although black holes resulting from the collision of gravitational radiation are most likely non-spherical, they are dynamic and quickly become spherical. The final state can approximately have a maximum angular momentum of a spherical mass whose radius is $R_H$, and whose angular velocity is $c/R_H$:

$$L_{max} \approx \tfrac{2}{5} M R_H^2 \left( \frac{c}{R_H} \right) = \tfrac{2}{5} M R_H c. \qquad (3.5)$$

Originally there was much debate as to whether or not EGR permits gravitational radiation. The theoretical existence of gravitational radiation in EGR won out, and is currently the orthodox view though there are still dissenters. This is supported by impressive indirect experimental evidence as mutually orbiting neutron stars have been observed to fall in toward each other, presumably as they lose energy due to gravitational radiation. Much experimental work is in progress for the direct observation of gravity waves, but none have been detected as yet. Yilmaz regards gravitational radiation as more natural in YGR than in EGR.

### 3.3 Recent progress and regress

The decision on Penrose's 1969 cosmic censor conjecture is still in a state of flux. As can be seen from the following examples, it remains an elusive question.

Shwetketu Virbhadra (1996) considers a static, axially symmetric, and asymptotically exact solution of the Einstein vacuum equations. He found that his solution has "directional nakedness" for some cases with possible implications for astrophysics. Virbhadra and Ellis (2002) fill a compelling void in both General Relativity and astrophysics by providing a gravitational lens equation for observationally testing the Penrose Cosmic Censorship Hypothesis with a cosmic telescope. Such gravitational lensing may enable empirical detection, and discrimination between singularities and black holes if in fact they can be found in our universe.

Singh (1998) argued that it appears that the cosmic censorship hypothesis may not hold in classical general relativity. He also looked at quantum processes that take place near a naked singularity, and their possible implication for observations. Joshi, Dadhich and Maartens (2002) investigate the key physical features that cause the development of a naked singularity, rather than a black hole, as the end-state of spherical gravitational collapse. They concluded that sufficiently strong shearing effects near the singularity delay the formation of a horizon.

In a 34 page paper (2000) Harada, Iguchi, and Nakao concluded that a naked singularity occurs in the gravitational collapse of an inhomogeneous dust ball from an initial density profile which is physically reasonable. They suggested that this phenomenon may provide a new candidate for a source of ultra high energy cosmic rays or a central engine of gamma ray bursts. In their 76 page review paper, Harada, Iguchi, and Nakao (2002) present examples of naked singularity formation However two years later, (2004) Harada and Nakao equivocate that it is still uncertain whether the cosmic censorship is true or not.

Many possible counterexamples to Penrose's cosmic censor conjecture have been presented over the past three decades, but none of them has proved to be sufficiently generic. Hertog, Horowitz, and Maeda (2004) conclude that cosmic censorship does not hold for certain reasonable matter theories. The asymptotically flat case is more subtle.

They suspect that potentials with a local Minkowski minimum may similarly lead to violations of cosmic censorship in asymptotically flat spacetimes, but do not have definitive results. They were quickly challenged by Alcubierre et al (2004) that there is a logical and physically plausible loophole in their argument, and that the numerical evidence in a related problem suggests that this loophole is in fact employed by physics

Frolov (2004) used string theory to investigate examples of cosmic censorship violation. He found no instances of the formation of naked singularities. Instead, his numerical calculations indicate that black holes form in the collapse, and show where the "no black hole" arguments break down.

**3.4 Discussion**

Two separate questions should be considered. Are there naked singularities in EGR; and are there naked singularities in nature? Since EGR certainly has singularities concealed by black hole horizons, we may broaden our inquiry of nature to include singularities of all kinds. The most reasonable approach is to reject singularities of any kind, altogether. Certainly continuum theories breakdown at singularities, or even below some threshold scale just as fluid mechanics breaks down at molecular scales. One approach to eliminating singularities, avoids them in the first place by means of a discrete rather than continuous formulation of space-time. Yet this also introduces conceptual difficulties such as what lies between space-time points. Somehow, the analogy that no integer lies between two consecutive integers -- although true -- is not altogether satisfying. The notion of a singularity in nature may be so abhorrent to science that it is readily dismissed, since science comes to a screeching halt at a singularity. Nevertheless, we are not the ones to decide how nature behaves, and such a possibility should, at least, be considered.

If we can decisively conclude that there are no singularities in nature, then we must consider how to eliminate them from our physical theories. Are there other ways besides considering only denumerable approaches such as that of Winterberg (2002), or those of Finkelstein (1996) in which nature is fundamentally discrete? Having no naked singularities in EGR would help. If all singularities were clothed inside black hole horizons, at least experimentalists would not have to cope with them. However, black holes evaporating by Hawking radiation decay down temporarily to a singularity, and EGR does admit of a singularity at the big bang.

Although the Big Bang is generally accepted, its initial singularity problem continues to be troubling. General relativity implies that at time zero, the universe was a point of zero volume with infinite density, temperature, etc. Gabriele Veneziano, one of the fathers of string theory, avoided this singularity by using string theory in which conventional point-like particles are replaced by one-dimensional strings having a very short length. Not only are the troublesome infinities eliminated, but string theory endows the universe with a past before the Big Bang. Previously we were told that there was no "before." Now we may view the evolution leading to the Big Bang in much the same way as the explosion or collapse of stars as detailed in a long, extensive report by Gasperini and Veneziano (2003).

It is tempting to use intuition with respect to the naked singularity paradox. Unsophisticated intuition may well lead one to conclude that there can be no naked singularities because there must be a region near a singularity where the escape velocity exceeds the velocity of light. But one should not leap to a conclusion. Even though many problems are less subtle than meets the eye -- because they are simpler than expected upon closer scrutiny -- this problem is more subtle than meets the eye. I am reminded of the question of where in a gravitationally closed orbit the gravitational force is equal and opposite to the inertial force $mv^2/r$? If the orbit is circular the two forces are equal and opposite everywhere. But what about an elliptical orbit? At first sight one might think that it is obvious that they are equal and opposite at only two points, the apogee and the perigee since there is a non-zero force component parallel to the orbiting body's instantaneous velocity everywhere except at those two points. However, since the apogee and perigee lie respectively outside and inside the corresponding circular orbit of the same total energy, the two forces must be unequal at those points as well. Another way to say it is that at perigee, the attractive force is less than $mv^2/r$ since the body is about to move outward. And at apogee, the attractive force is greater than $mv^2/r$ since the body is about to move inward. So we go from the two forces being equal everywhere for a circular orbit, to the two forces being unequal everywhere for an elliptical orbit. A small change can make a big difference as happens in the Yilmaz modification of EGR which eliminates black holes altogether.

It would be nice to be able to tackle the naked singularity question without the need for computer simulations. An example of a problem that is extremely difficult to solve by computer simulation, but can easily be solved analytically comes to mind as an illustration. Consider a charged spherical shell that is expanding and contracting radially. Each element of charge radiates, yet there is complete destructive interference with no net radiation whatsoever. This would be exceedingly hard to prove by computer calculation. Yet the analytic approach is easy and exact. The center of charge does not move, so that a distant observer sees only a static electric field. Radiation requires a time-varying electric field, E. Or an equivalent argument is that there is no net current and no magnetic field H because of spherical symmetry, and hence the Poynting vector $\vec{E} \times \vec{H}$ is zero. Radiation requires a non-zero Poynting vector.

## 4 Black Hole Lost Information Paradox (LIP)

Hawking (1976) shocked the physics community with his claim that information is lost in a black hole -- not just stored to be later released by Hawking radiation, but completely destroyed. This conclusion is not consistent with Hawking's approach of superimposing quantum field theory on classical EGR to derive Hawking radiation. Although he was challenged by many luminaries, Hawking not only held fast, but he was cocksure that he was right and all of them were wrong. He did not equivocate in his position that when information in any form falls into a black hole it completely vanishes from our universe and can never be retrieved -- not even in scrambled form.

Prior to Hawking radiation one would expect information -- the specific forms that energy takes -- to be forever lost to distant observers as nothing was expected to ever come out of a black hole. However, observers could in principle go inside a black hole and detect the information that went in -- even though they could never communicate their findings. But if Hawking radiation exists, this changes the nature of the problem, since the black hole can evaporate away completely.

Hawking's conclusion that the information that entered a black hole can be forever lost comes directly from his consideration that black hole radiation is featureless thermal black body radiation, which loses information about its source. His position implies that black hole evolution is non-unitary, violating one of the most basic principles of quantum theory. It is not necessary to attribute this to his position that the radiation does not originate from within a black hole, but comes from the vicinity outside it due to particle anti-particle creation. If it only appears to come from within, then it would not reflect what is inside as the hole evaporates away. He argued that as the black hole radiates, it will eventually completely evaporate away. The resulting radiation state would be precisely thermal so there would be no way to retrieve the initial state. On the other hand, time-reversal symmetry, classical and quantum physics are violated if the information is lost.

Quantum theory allows for an infinite number of representations, depending on which complete set of base functions are chosen for each representation. One representation in a given base set may be transformed in terms of another by a similarity transformation or a unitary transformation. For example, a unitary transformation between the representations of incoming and outgoing information implies the conservation of information. A unitary transformation is more restrictive with properties not shared by a general similarity transformation. For example, the Hermitean is invariant in unitarity, but not in similarity transformations. [A Hermitean operator is one in which each matrix element is equal to the complex conjugate of the corresponding element of the transposed matrix.] Unitarity or its violation is central to most papers that deal with the LIP. For those who would like to be refreshed, let us summarize four important properties following the more detailed excellent treatment by Bohm (1958). In a unitary transformation:

1) The normalization of an arbitrary wave function is left unchanged.
2) The transformation of wave functions that were originally orthogonal, remain orthogonal.
3) The relationships between transformed operators are the same as those between the corresponding untransformed operators.
4) The transformation does not change the eigenvalues of a matrix.

## 4.1 Preskill

John Preskill (1992) is a good starting point for understanding alternative solutions to the LIP, because he summarizes the possible ways he and others have tried to resolve the LIP since it was introduced by Hawking (1976):

1) Can the information come out with the Hawking radiation?
2) Can the information be retained by a stable black hole remnant?
3) Can all of the information come out at the end?
4) Can the information be encoded in quantum hair?
5) Can the information escape to a baby universe?

Preskill finds that for him, the fifth possibility is the most satisfying resolution of the LIP. In this view, quantum gravity effects prevent the collapsing body from producing a true singularity inside the black hole. Instead a closed "baby universe" is nucleated. This new universe retains the lost information, which is unobtainable or even detectable by us. However, it would be accessible to a "superobserver" who unlike ourselves, can observe both our universe and the baby universe. Although such baby universes have been appealing to Preskill, Hawking, and others, they have an aura of science fiction about them; and have been largely abandoned.

With respect to the first possibility that the information comes out in the Hawking radiation, Preskill presents an analogy that makes an apt distinction between information lost in practice and information lost in principle by considering an encyclopedia that falls into the sun:

> But we don't really believe that the information about the initial quantum state has been lost in principle. Even as the encyclopedia burns beyond recognition, all of the information that it carried presumably becomes stored in subtle and intricate correlations among the radiation quanta emitted by the sun, or correlations of the emitted quanta with the internal state of the sun. Information is lost in practice because we are unable to keep track of all these correlations.

Preskill finds that all the above approaches have sobering shortcomings. This leads him to conclude that the LIP "may well presage a revolution in fundamental physics."

### 4.2 Susskind et al

Leonard Susskind, one of the founders of string theory, was among the first to challenge Hawking informally as soon as he made his pronouncement. Susskind seems to be the first to have realized that string theory could shed light on black holes and the LIP since it is a theory grounded in gravity. Susskind and Thoriacius (1992) presented a formal argument that not all the information is lost or drained out of the incoming field. They argued that if the infalling information is not stably stored, then the only way to have a unitary quantum theory is to find a complete correlation between the quantum state of the infalling matter collapsing into the black hole, and final quantum state of the emitted Hawking radiation. They felt it imperative to demonstrate a mechanism for transferring the imploding information to the outgoing radiation.

Susskind (1993) pursued information loss in terms of string theory, and the principle of "black hole complementarity" which deals with virtual inconsistencies. For him, the LIP boils down to the localization of information and how it is perceived by different observers. He claims that "black hole complementarity" makes copacetic the following apparently disparate assumptions:

> 1) To a freely falling observer, matter falling toward a black hole encounters nothing out of the ordinary upon crossing the horizon. All quantum information contained in the initial matter passes freely to the interior of the black hole.

> 2) To an observer outside the black hole, matter, upon reaching the "stretched horizon", is disrupted and emitted as thermalized radiation before crossing the horizon. All quantum information contained in the initial matter is found in the emitted radiation.

We must bear in mind that these two statements are what need to be proven, and are still assumptions -- albeit presumably consistent assumptions though they appear to be inconsistent.  The "stretched horizon" can be thought of as a membrane which lies just outside the black hole event horizon.

Susskind concludes that the information carried by a freely falling particle is localized for a co-falling observer as they fall through the horizon.  However, a distant observer sees a totally different story.
For a distant observer, the particle and its information content expands to fill the entire black hole area in a time

$$t \sim \frac{G^2 M^3}{\hbar c^4} . \qquad (4.1)$$

However this "information retention time" is just the time that it takes the entire black hole to evaporate away by Hawking radiation.  This compromises the view that the information is retained, i.e. not lost.  As we shall see in Sec. 4.4.2, this problem is avoided in gravitational tunneling radiation.

Susskind and Thoriacius (1994)  further explored the LIP by considering several gedanken experiments.  They found that if information leaves the black hole there is no hint of it left inside the black hole.  That is, observers who sample Hawking radiation before entering the black hole, do not discover any duplicate information inside the black hole before hitting the singularity.  This finding seems to be at odds with the dictum that a freely falling observer can't even tell when he passes through the horizon of a massive black hole.  There is a slight increase in tidal force, which happens even outside the black hole.  The dictum says that the only experiment he can perform in his falling frame to tell him when he has crossed the horizon is that he cannot reverse the direction of his motion.  If Hawking radiation stop for him inside the horizon,  can he detect a decrease in information as he enters?   Perhaps the dictum needs to be modified.

For Susskind and Thoriacius (1994) one of the most exciting undecided questions in theoretical physics is "whether the unitary evolution of states in quantum theory is violated by gravitational effects." Posed in terms of the LIP, the quandary is that almost no information about the initial state of the imploding matter emanates in the Hawking radiation that leaves at faster and faster rates until there is nothing left of the black hole.  And although properties are attributed to the vacuum, no one has yet suggested that the lost information resides in empty space.

They point out that many everyday processes have much in common with the LIP in that organized energy and information is absorbed, thermalized, and radiated away. Yet in principle in these processes, information can be retrieved. They do not offer a resolution of the LIP. Instead they conclude, "the information paradox can only be precisely formulated in the context of a complete theory of quantum gravity, and that the issue of information loss cannot be definitively settled without such a theory." In other words, the LIP cannot be well enough formulated at present to resolve it. They also show that the question of baryon number violation is related to the LIP. It is their "view that black hole complementarity is not derivable from a conventional local quantum field theory." I will present an argument in Sec. 4.4.2 to show that black hole complementarity as related to the LIP and the LIP are both challenged by gravitational tunneling radiation.

In a cogent paper, Susskind and Uglum (1996) demonstrate that after information enters a black hole, it cannot possibly get out. Therefore they appear to agree with Hawking that the information is lost to outside observers. However in a surprising twist, they point out that their apparently undeniable argument for information loss includes a critical assumption that their low energy theory is a local field theory. Since this assumption is not at all self-evident, the door is left open. If it can be shown to be wrong, this does not prove that information is conserved -- it only nullifies the proof that information is lost in a black hole.

To understand how information is coded and potentially retrieved in ordinary systems, they considered a gedanken experiment suggested by Sidney Coleman and solved by Don Page:

> Illuminate a piece of coal with a sequence of pulses from a laser beam. The coal radiates the energy absorbed from the laser in the form of thermal radiation. This continues until the coal cools down to its ground state. Since we know that the evolution of this process is unitary, the information coded in the sequence of pulses must still be present. Since the coal has cooled back down to its ground state, it contains no remnant information. So the information must be contained in the radiation field.

The LIP applies to all matter that falls into the black hole, from the matter that originally created the black hole to any matter that falls in subsequently. Their focus on the laser and coal analogy covers only the latter half of the LIP -- albeit an important half -- and overlooks the much more difficult half. It applies only to recovering the laser supplied information, and is directed neither at recovering information about the creation of the coal, nor even to its history after creation and prior to laser excitation. Creation of a black hole is a much more irreversible process than subsequent falling in of matter, since it clearly involves the creation of much more entropy.  Granted that recovering the laser-coded information is difficult enough by itself, we'll look at how this is accomplished in Sec. 4.3.

### 4.3  't Hooft et al

Stephens, 't Hooft, and Whiting (1994) assume that a black hole preserves quantum coherence.  This is an assumption for which there is neither unequivocal corroboration nor indisputable proof for the antithesis.  One of the problems in dealing with the black hole LIP is that the vacuum state is not well defined in terms of a stable state of minimum energy, i.e. one that is bounded from below.  Thus they have to pick subjective, though reasonable, ways of avoiding vacuum state pitfalls.

Within the framework of a model, they demonstrate that the formation of a black hole obeys unitarity, despite the fact that the spectrum of emitted particles is approximately thermal.   They show that "super observers" cannot exist that detect both outgoing Hawking radiation and in-falling matter.  In accord with Susskind et al, they find that infalling and distant observers can only perceive different physical systems.  For them this obviates inconsistencies between in- and far-observers because they can never compare data.  They conclude that, "They only believe that they have a promising avenue of investigation for the future with which to examine this question."   And it is not their intent, "to offer yet a definitive answer concerning whether black hole formation or decay leads inevitably to the loss of quantum coherence."

It is tempting to surmise that Gerard 't Hooft's work on the LIP led him to his ground-breaking concept that dissipation of information may lead to quantum mechanics from a more fundamental deterministic theory (1999, 2000, 2002).  His struggle with the vacuum state carries over from the LIP in that he finds that it is still "difficult to obtain a Hamiltonian that is bounded from below, and whose ground state is a vacuum that can exhibit complicated vacuum fluctuations, as in the real world."  He enunciates  a need for new physics at the Planck scale.  It is interesting to note some similarities in 't Hooft's approach with that of Winterberg (2002)  and Rivas (2001).

### 4.4  Additional opposition to the loss of information

### 4.4.1 *Page*

Don Page (1993 a) presented a definitive analysis to Coleman's question described in Sec. 4.1 as to how information is coded and potentially retrieved in ordinary systems such as a lump of coal zapped by a laser beam. The coal is initially in a ground state at 0 K, and hence at 0 entropy by the Nernst heat theorem. Non-random laser illumination with long-duration, long-distance correlations between photons puts the coal in an excited state of high entropy -- leaving the radiation field in a state of low entropy at the end of this process. When the coal stops its thermal (random) radiating and goes back to its ground state (0 K in this gedanken experiment) its entropy goes back to 0, and hence the coal contains no remnant information. By unitarity all the original information must be returned to the radiation field, which is now in a state of high entropy.

Page demonstrated that at least half of the combined coal-radiation system must be sampled before even one bit of information can be retrieved. Therefore the information is contained in long-distance correlations, or it could be retrieved by scrutiny of moderate-sized parts of the system. The analysis is motivated by the LIP, rather than a mundane system like a piece of hot coal. The analogy is that the information may be stored in long-time correlations in the Hawking radiation. To push the analogy all the way, to recover one bit of the information one may need to examine at least half of the combined black hole-radiation system. But the black hole does not stand by idly while the system is being sampled. As a result much information is lost due to Hawking evaporation, as the system is being scrutinized to find the information.

Despite the fact that all black holes can be characterized by a few variables like mass, charge, and angular momentum large black holes have many degrees of freedom for storing energy, just as do coal or other ordinary macroscopic objects. The ability to determine their information content as related to whether such systems are in a single state or mixed state would be a daunting, time-consuming task. The coal can patiently stand by for inspection, but not a black hole that is decaying by Hawking radiation. Yet it is not obvious whether a significant fraction of a sizable black hole could be examined.

For a black hole of mass M greater than a Planck mass, the evaporation rate due to Hawking radiation is

$$\frac{d\left(Mc^2\right)}{dt} = -P_H = -\left[\frac{\hbar^4 c^8 \sigma}{16\pi^3 k^4 G^2}\right]\frac{1}{M^2}, \qquad (4.2)$$

where k is the Boltzmann constant, and $\sigma$ is the Stefan-Boltzmann constant. One may obtain $\sigma$ by integrating the Planck distribution over all frequencies, giving

$$\sigma = \left\{ \frac{\pi^2 k^4}{60 \hbar^3 c^2} \right\}. \tag{4.3}$$

We obtain the black hole lifetime by integrating eq. (4.2) and substituting eq. (4.3) for $\sigma$:

$$t = \left[\frac{16\pi^2 k^4 G^2}{3\hbar^4 c^6}\right]\left\{\frac{1}{\sigma}\right\}M^3 = \left[\frac{16\pi^2 k^4 G^2}{3\hbar^4 c^6}\right]\left\{\frac{60\hbar^3 c^2}{\pi^2 k^4}\right\}M^3 = \left[\frac{320\pi G^2}{\hbar c^4}\right]M^3. \tag{4.4}$$

Page (1993 b) maintained that if information comes out in black hole radiation, it does so at an initial rate so low that it couldn't be found theoretically or experimentally. It wouldn't show up in an order-by-order perturbative analysis, and too many measurements would be needed to find it.

Susskind (1993), and Susskind et al (1994 and 1996) argued that because the information leaving the black hole as thermal radiation is coded in long-time, long-distance correlations between the photons one must wait for a time of the order of half the lifetime of the black hole to retrieve even one bit of the initial information. In these papers, the argument is presented only in order of magnitude form yielding a waiting or information retention time $t \sim M^3$, and it is pointed out that this is also the dependency for the lifetime of the black hole. In terms of fundamental constants $t \sim \frac{G^2 M^3}{\hbar c^4}$. Since there is a factor of $320\pi$ in eq. (4.4) for the black hole lifetime, these two times may differ by as much as three orders of magnitude, but it is not possible to tell from the analysis shown. As appealing as this limitation of lifetime and retention time analysis is, it may not be airtight as shown in Sec. 4.4.2.

### 4.4.2 *Rabinowitz*

Mario Rabinowitz (2003, 2001a) was the first to challenge the LIP based on a physically reasonable alternative to Hawking radiation that could also avoid several other quandaries, and give insight into many other phenomena. Gravitational tunneling radiation (GTR) is an alternative to the thermal blackbody-like Hawking radiation. A consequence of GTR is that it resolves the lost information paradox, since the radiation comes from inside the hole and carries attenuated but undistorted information from within. Since it is a tunneling process and not an information voiding Planckian black body radiation distribution, it can carry information related to the formation of a black hole, and avoid the information paradox associated with Hawking radiation.

GTR gives a different perspective with respect to the lifetime and retention time analysis given by Susskind et al. in Sec. 4.4.1. The disparity for these two times could be much, much greater than three orders of magnitude if, instead of evaporating by Hawking radiation, the black hole evaporates or decays by gravitational tunneling radiation (Rabinowitz, 1999 a,b, 2003). The "information retention time" is just the time that it takes the entire black hole to evaporate away. They argue that if it evaporates by Hawking radiation the black hole lifetime is comparable to the sampling time. This compromises the view that the information is retained, i.e. not lost. However the much longer lifetime of black holes by gravitational tunneling radiation avoids this problem altogether.

For a black hole of mass M, the radiated power due to GTR is

$$P_R = 60 \langle e^{-2\Delta\gamma} \rangle \left[ \frac{\hbar^4 c^8 \sigma}{16\pi^3 k^4 G^2} \right] \frac{1}{M^2} = \frac{-d(Mc^2)}{dt}, \qquad (4.5)$$

(Rabinowitz, 1999 a, b) where $\langle e^{-2\Delta\gamma} \rangle$ is the average tunneling probability between the black hole and a second body. The process of GTR is analogous to electrical field emission. As shown, for GTR the "tunneling probability or penetration coefficient" is approximately equal to the "transmission probability" or "transmission coefficient." Solving eq. (4.5) for the black hole GTR lifetime

$$t_{GTR} = \left[ \frac{16\pi G^2}{3\hbar c^4} \right] \frac{M^3}{\langle e^{-2\Delta\gamma} \rangle}. \qquad (4.6)$$

Since $\langle e^{-2\Delta\gamma} \rangle$ can be extremely small, $t_{GTR} >>> t$. Thus in GTR, a black hole can live more than long enough to sample all the stored information, challenging the ingeniously conceived black hole complementarity. In fact GTR questions the entire lost information paradox.

**4.4.3 Jacobson**

Ted Jacobson (1998) argues that even though ordinary quantum mechanics is unitary, the LIP suggests that quantum gravity is not unitary. Two types of information are present: the state of the matter that falls into the black hole, and the correlations between the radiation quanta. He allows that both kinds are completely destroyed when the entire black hole decays away.

As previously discussed, in one point of view unitarity hinges on the singularity at the center of the black hole. The information that falls into the singularity might:

    a) be destroyed; or

    b) pass on to a baby universe born at the singularity; or

    c) leak out of the black hole in a locality-violating process.

Jacobson focuses on a string theoretic approach which leads him to conclude that the "singularity at the origin is irrelevant" to the LIP.

### 4.4.4 *Srednicki*

Mark Srednicki (2002) considers the detection of information radiated from a black hole. His resolution of the LIP is that in one formulation of physical theory, information is preserved and macroscopic causality is violated. However, in another formulation, causality is preserved and pure states evolve into mixed states causing a loss of information. In a kind of black hole complementarity, he concludes that it is not possible to perform an experiment that would distinguish these two descriptions.

### 4.5 Hawking recants

The views of many scientists on the LIP have been presented . It is clear that by and large they say different things than the sound bites in the press that are largely based on bets made before the principals had given much thought to the principles. Judging from the press, not only was Hawking's notion initially rejected by other scientists, that in principle information is irretrievably lost in a black hole, but most have continued to reject it. However judging from the scientific literature, most have moved towards Hawking's original position and almost embraced it. So it came as quite a surprise to find that Hawking has recanted the LIP to avoid embarrassing contradictions with classical and quantum physics.

Hawking now evades the paradox altogether by morphing the definition of black holes to avoid the theoretical embarrassment of contradicting the principle of conservation of information.  If information is lost, then physical processes are not predictable, and worse yet, not even retrodictable.  Not only would time reversibility be challenged, but so would the conservation of mass/energy.  In 2004, at the 17th International Conference of General Relativity and Gravitation, Hawking announced that all is not lost because  he was wrong about a key argument he put forward 30 years ago, and he has now solved the LIP. His argument sounds somewhat like Schroedinger's cat paradox that according to quantum mechanics the cat is both half dead and half alive. For Hawking, because of quantum uncertainty one can never be sure from a distance that a black hole has really formed.  This has been long recognized, since time gets frozen as a star implodes to become a black hole due to gravitational red shift as discussed in Secs. 2.2 and 2.3.  That is why the name "frozen star" was in vogue long before the name black hole was popularized.

And so Hawking's resolution hangs on the idea that for a distant observer there is no way to distinguish between a real and a virtual black hole. He claims that quantum fluctuations prevent a true event horizon from forming.  He asserts that information loss is only virtual for intermediate times, and that at long times the information is recoverable as the black hole vanishes away.  He concludes that for a virtual black hole, the apparent event horizon can unravel like a ball of string releasing non-thermal (non-random) radiation that  carries information.  This is somewhat like some of the arguments in earlier parts of Sec. 4 that do not invoke unraveling of a virtual black hole.  However, his virtual black holes emit radiation for extended periods, and eventually open up to reveal the remaining information within them.  Hawking adds that although he used to think so,  no baby universe is born inside the black hole.  The jury of scientific consensus is still out on his new solution in part because to date he has only presented an outline of this new view.  A published version of his new theory with detailed analysis that supports his general conclusions is not yet available, as of the date of submission of this chapter.

## 4.6  Discussion

Information is often lost in the  real-world performance of experiments.  However a basic tenet of  both classical and quantum mechanics is that it cannot be lost in principle. Information is stored both in an object that is destroyed *and* in the process of creating it. If an encyclopedia fell into a black hole, the printing process which produced it would leave abundant information about its contents outside the BH, and similarly for other objects. I think important information is lost in another sense. The otherwise separate locations of all objects falling into a black hole are reduced to one location which is quite indeterminate in the case of a large massive black hole, or very precise in the case of a LBH. When they are both moving at the same velocity, it is interesting that *this is just the inverse of the Uncertainty Principle* which yields more precise localization in the more massive and hence higher momentum black hole.

As Preskill (1992) points out, the LIP claims that "information is actually destroyed in principle." This follows from the nature of Hawking radiation. GTR is founded on a more sound basis, and since it does not lead to such a radical conclusion, it is more likely to be correct. At what point should a paradox be judged to be a self-contradiction? Does the lost information paradox hint at a flaw in EGR; or is the flaw more narrowly limited to Hawking radiation? The orthodox view is that classically, black holes swallow up information to be forever lost, and that quantum mechanics comes to the rescue with Hawking radiation. As we have seen, this rescue may only be partial, if at all. It is ironic that classically in EGR the information may be lost, and that quantum physics -- which inherently limits information -- prevents the total loss of information. Classically in NG, it is possible for a second body to lower the potential energy barrier and produce an analog of Schottky emission from Newtonian black holes that loses no information. So the dichotomy is really not between classical and quantum physics, but between EGR and the quantum realm.

It is generally accepted that, aside from possible correlations between radiated quanta, black hole radiation is independent of the time-history of black hole formation. So from one point of view, for theoretical purposes, a black hole may be considered that has existed for an infinite length of time. In this case, an observer would have an indefinite amount of time to inspect it for stored information. Conclusions about the radiation of such a static everlasting black hole should also be valid for large and small black holes, despite the virtual inconsistency that any black hole would have evaporated away after an infinite time. From another point of view, a small black hole can easily vanish in a very short time leaving nothing behind but radiation. Thus the remaining radiation is the entire final system. This in itself is not so bad if the final state were a pure state such as with gravitational tunneling radiation (GTR) which does not thermalize the radiation. However, Hawking radiation is a mixed state because it is thermal. So either an initially pure quantum state that collapses into a black hole becomes a mixed state by becoming total Hawking radiation (which has never been detected in over a quarter of a century), thus violating quantum mechanics, or perhaps Hawking radiation does not exist.

## 5 Why Hawking Radiation May Not Exist

In Hawking's model of black hole radiation, quantum field theory is superimposed on curved space-time, and gravity is described classically according to Einstein's general relativity. It is a semiclassical approach in which only the matter fields are quantized, and black hole evaporation is driven by quantum fluctuations of these fields. Hartle and Hawking (1976) and Gibbons and Hawking (1977) claim that the pair-creation model of Hawking radiation is equivalent to considering "... the positive and negative energy particles ... as being the same particle which tunnels out from the black hole..." Their tunneling is quite different from and has different consequences than GTR (Rabinowitz, 1998, 1999 a,b, 2001a, 2003). Aside from problems of dealing with infinities, they require a very high degree of correlation between incoming negative energy radiation and outgoing positive energy radiation that is likely unphysical. After over a quarter of a century Hawking radiation has neither been detected experimentally anywhere in the universe, nor has intense theoretical effort succeeded in predicting the time history of black hole decay.

### 5.1 Belinski: Hawking radiation does not exist

Vladimir Belinski (1995), a noted authority in the field of general relativity, unequivocally concludes:

> the effect [Hawking radiation] does not exist.

He comes to the same conclusion regarding Unruh (1976)-Davies (1975) radiation. He argues against Hawking radiation due to the infinite frequency of wave modes at the black hole horizon, and that the effect is merely an artifact resulting from an inadequate treatment of singularities. It is noteworthy that Unruh (1976) maintains that:

> an accelerated detector even in flat spacetime will detect particles in the vacuum. ... a geodesic detector near the horizon will not see the Hawking flux of particles.

Belinski probes deeper than this, since the fact that a derivation is invalid does not disprove the existence of an effect. He goes on to set up the problem with what in his terms are proper finite wave modes. He then concludes that no particle creation can occur.  As explained in Sec. 6, the Rabinowitz Gravitational Tunneling Radiation model (1999 a,b, 2001a, 2003) involves no infinite frequency wave modes, and does not invoke the creation of particle-antiparticle pairs to produce black hole radiation.

Belinski determines that one reason the Hawking and Unruh effects are mathematical artifacts is that they violate a principle of quantum theory:

> which does not permit the physical particle wave functions to have singularities at those space-time points at which the external field is regular and where there are no sources.

This happens at the horizon in their derivations despite the fact that all horizon points are regular and free of any sources.  There have been attempts to resolve this problem with a high frequency cutoff.  However, this is antithetical to relativity theory, and is more questionable than the problem being rectified.

Belinski objects to what is called the "backreaction on the metric" which is missing in Hawking's derivation.  This objection was raised by many researchers, and from Belinski's perspective has not really been laid to rest. Belinski shows that if done properly there is no real particle-antiparticle creation because of "The inability of the particle to cross the barrier between the two Dirac seas ...."  Infinities are manipulated in ways that are not justifiable.  Since there is no experimental verification of Hawking or Unruh radiation, they are not experimentally justified.

## 5.2  Wald calls Hawking radiation into question

Robert Wald (1992) makes incisive observations that can be interpreted as being critical of Hawking radiation, but not as critical as those of Belinski (1995).  He points out that although Hawking has found a sophisticated mathematical way of associating a thermal state to Schwarzschild spacetime, actual physical thermal emission has not been derived.  It has not been shown that the thermal state will arise by either a plausible or implausible process, even if the space-time were shown to be physically relevant.  Although I agree with most of Wald's points, I would say if an authentic thermal state has been achieved in Schwarzchild spacetime, it follow that there is radiation.  However, if the thermal state does not have a true thermodynamic temperature or is just a mathematical artifact, then there may not be radiation.

Wald does allow that thermal emission is a physical process in Kerr space-time for a rotating black hole.  However this seems closely akin to Zeldovich superradiance from a black hole.  Zel'dovich (1971) was the first to propose a model of radiation from a black hole:

> The rotating body [black hole] produces spontaneous pair production [and] in the case when the body can absorb one of the particles, ... the other (anti)particle goes off to infinity and carries away energy and angular momentum.

This is quite similar to the model proposed three years later by Hawking (1974, 1975) for radiation from non-rotating black holes.

Wald (1977) was one of the first to raise the "backreaction on the metric" objection.  Unruh and Wald (1984) raise and claim to resolve "Several paradoxical aspects of this process related to causality and energy conservation... ."  The recent consideration of related anomalies (Balbinot, Fabbri, and Shapiro, 1999) is testimony that the backreaction problem is still in need of better understanding.  They say, "Notwithstanding decades of intensive studies, the evolution and fate of an evaporating black hole are still unknown."

**5.3  Additional opposition to Hawking radiation**

Many reputable scientists questioned the validity of the Hawking model (1974, 1975) not long after its introduction.  Belinski (1995) is not the only one to question the existence of Hawking radiation in recent times.  His is a compelling and recent challenge.  Some of the other challenges have been both less manifest and less direct.  De Sabbata and Sivaram (1992) suggest that "Thus one may observe the decay [Hawking radiation] only if one makes an infinite succession of measurements. So in a sense one may never be able to observe the Hawking effect." Balbinot (1986) concluded that highly charged black holes do not radiate.  He concludes that "For an extreme Reissner-Nordstrom black hole ... there is no Hawking evaporation."  As a black hole becomes more and more charged, the Hawking radiation decreases until there is none.  The maximum charge it can hold is when the electrical potential energy equals
twice the gravitational potential energy:

$$z^2 e^2 / 4\pi\varepsilon r = 2GM^2 / r \Rightarrow z = (M/e)[8\pi\varepsilon G]^{1/2} = 7.6 \times 10^8 \, M_{kg}.$$
(5.1)

So a $10^{-3}$ kg  (1gm) LBH could hold up to a maximum of $7.6 \times 10^5$ net electron or proton charges.  A LBH could get charged in going through the atmosphere and lightning clouds to approach becoming an extreme

Reissner-Nordstrom black hole with significantly reduced Hawking radiation, but with negligible reduction of Gravitational Tunneling Radiation (Rabinowitz, 1999 a,b, 2001a). This may also have happened in the earlier more dense universe.

**5.4 Effects of Compacted Higher Dimensions**

Another approach assumes the correctness of the Hawking model, but analyzes the effects of additional compact higher dimensions on the attenuation of this radiation. Argyres et al (1998) conclude that the properties of LBH are greatly altered and LBH radiation is considerably attenuated from that of Hawking's prediction. Their LBH are trapped by branes (which may be thought of as a vibrating membrane) so essentially only gravitons can get through the brane. For them, not only is the radiation rate as much as a factor of $10^{38}$ lower, but it also differs in being almost entirely gravitons.

# 6 Gravitational Tunneling Radiation (GTR): Alternative or Adjunct to Hawking Radiation

The paradoxes and inconsistencies engendered by Hawking radiation trouble me. Furthermore, Hawking's derivation (1974, 1975) has a number of bothersome aspects as discussed in Sec. 5. Not the least of which is that it deals with space-time capriciously adjacent to the black hole horizon, and thus includes arbitrarily large energies, and an inordinately large gravitational redshift. These are infinite at the black hole horizon.

The main difference between the other critics of Hawking radiation and my criticism is that they do not offer an alternative mode of black hole radiation to replace it or even act as an adjunct mode, whereas my GTR model does. GTR may not be a fully accurate or only model of radiation from a black hole. It is presented as a proposal for investigation. As either an alternative or adjunct mode of radiation, GTR can void the lost information paradox and provide a possible explanation for the accelerated expansion of the universe (Rabinowitz, 1999b, 2003). As an alternative to Hawking radiation, GTR fills an important gap if:

a) Hawking radiation is not physically real.

b) Hawking radiation is absent as in an extreme Reissner-Nordstrom charged black hole.

To my knowledge, none of the proposed solutions to the LIP that invoke at least partial retrieval of the information by means of the emitted Hawking radiation deal with the case when there can be no Hawking radiation. As discussed in Sec. 5.3, when a black hole becomes more and more charged, the Hawking radiation eventually ceases.

Furthermore, the reduced radiation of GTR (cf. Sec. 4.4.2) compared with Hawking radiation allows LBH to be candidates for dark matter, i.e. the 95% of the missing mass of the universe (Rabinowitz, 1999 a, b, 2005). For Hawking that many LBH would fry the universe. That is why he concludes that his LBH can't be more than one-millionth of the mass of the universe.

## 7 Paradoxes of Black Hole Entropy

Closely related to the lost information paradox is the paradox of black hole entropy. The apparent incongruity of the irreversible entropic arrow of time with our time-symmetric dynamic laws is an intriguing, well known, and yet unanswered question explored with crystal clarity by Penrose (1979). An equally intriguing question is: *What happens to entropy that is put inside a black hole? Does this entropy disappear and the entropy of the universe go down?* Jacob Bekenstein (1972, 1973, 1974) conceived of black hole entropy in response to this conundrum posed to him by John Wheeler. His synthesis appears to have uncovered a deep-rooted connection between thermodynamics, gravity, and quantum physics.

The accelerated expansion of the universe notwithstanding, the laws governing gravity are mature, are closely confirmed by experiment, and appear to be well understood. On the other hand, the thermodynamics and in particular the entropy of general gravitational systems leaves much to be desired. Their ability to self-organize seems to be at odds with the second law of thermodynamics in that the apparent entropy of the system decreases. Perhaps if we look closely enough, or properly modify the entropic concept, the overall entropy will be found to increase. In any system, there is a small probability that the entropy will decrease. Is the ability of the universe to self-organize nothing more than a statistical fluctuation, not apparent from our vantage point? Our view of entropy may have to be as pliable as our view of the conservation of energy. There are also those who argue that the principle that entropy can only increase cannot be applied to the entire universe. For them such an extension is an unwarranted extrapolation founded neither on fact based on experience, nor on indisputable principle (Zucker, 1999).

Though black holes are far from being well understood, the general orthodox view is that black hole entropy is on a firm and well understood basis. Nevertheless, we should be vigilant, lest we become too complacent to challenge common consensus. Black hole entropy does not represent a quantum limitation to entropy, but rather a quantum extension of the entropy concept. Black hole entropy may prove to be as malleable as the conservation of energy (Rabinowitz, 2005). Whenever energy conservation was challenged, a way was found to preserve it.

**7.1 The concept of black hole temperature**

Temperature is not a well-defined concept for a black hole. Temperature is fundamentally a statistical concept requiring many bodies. If the distribution of energies is broad, the temperature is high. If the distribution of energies is narrow, the temperature is low. Temperature is basically a measure of the half-width of the distribution. For the sake of illustration and concreteness, let us consider the concept of temperature in the context of a Maxwell-Boltzmann distribution (MBD).

Randomness of the velocity vector, $\vec{v} = \vec{v}_x + \vec{v}_y + \vec{v}_z$ in velocity phase space, for particles of mass m, is the primary assumption leading to the MBD:

$$N(\vec{v}) = \frac{d^3 N(v_x, v_y, v_z)}{dv_x dv_y dv_z} = \left(\frac{N\beta}{\pi}\right)^{3/2} \exp(-mv^2/2kT) \ . \qquad (7.1)$$

The connection to temperature T,

$$\beta = \sqrt{\frac{m}{2kT}} \qquad (7.2)$$

is made only after additional assumptions are introduced, such as equipartition of energy and the form of the potential energy, that may not apply to a black hole. For a black hole it not clear what the distribution is. If states are confined to a Planck area or volume, it is not clear what the states are.

The MBD of eq. (7.1) may generally be used, and the same applies to other distributions. However, for an ensemble of particles that are exchanging kinetic and potential energy near (or in) a black hole, the connection to temperature is not at all as straightforward as is usually assumed. A further complication arises because the flux velocity v needs to be modified to take into account the exchange of kinetic and potential energies as the particles make their journey from a black hole to a distant observer.

Conventionally, one may relate the mean kinetic energy to the temperature for point particles thus

$$\left\langle \tfrac{1}{2}mv^2 \right\rangle \equiv \int_0^\infty v^2 N(v)dv = \tfrac{3}{2}kT \ . \tag{7.3}$$

This is the way that an effective temperature is usually assigned to a system of unknown distribution function. What is often forgotten is that this assignment presupposes a distribution function. This assignment of T as a real temperature has an element of arbitrariness to it. Although the often used statistical mechanics assignment of $\tfrac{1}{2}kT$ per degree of freedom is valid for many systems, it is not generally valid. For example it does not apply to the vertical motion of a gas in a gravitational field, and as will be shown in Sec. 8.5 it is not valid for thermionically emitted electrons. It is also invalid for the energy associated with molecular rotation, vibration, and electronic excitation. This is because of the quantized nature of these energies which can take on only certain discrete values, and are not a continuous function of some coordinate. That is why their energy cannot be expressed as a simple linear function of the temperature.

    For all of the above reasons a real thermodynamic temperature may not be well justified for any size black hole. There may be a distribution function during the collapse process, but unless there is a dynamical process at the black hole horizon such as may be related to its radiation, the temperature may only be virtual. For an LBH there is a further blemish because the system is far from thermodynamic equilibrium since it is radiating energy at a high rate. In their paper on the laws of thermodynamics of black holes, Bardeen, Carter, and Hawking (1973) very specifically said that the temperature they assigned was not real, but just an effective temperature. **They also were very careful to add that for them the real temperature of a black hole is zero.** The latter was the basis on which Hawking attacked Bekenstein's concept of black hole entropy. Hawking argued that if a black hole has entropy this implies that a black hole has a real temperature greater than 0 and that we all know that the temperature of a black hole must be 0. If a black hole has a real temperature (is thermalized), then it must radiate, but this was **before** Hawking radiation. If black hole entropy were configurational as it is in information theory, then in my opinion there would be no reason to assign a physical temperature to it. This all changed after it was realized that black holes can radiate.

    A simple approach which is both heuristic and quantum mechanical in considering a quantum particle inside the black hole can yield an effective temperature, T, of the particles emitted by GTR from a black hole. In GTR the emitted particles do not undergo a gravitational red shift in tunneling through the barrier. Thus the temperature and

average energy of the particles emitted by tunneling is the same on either side of the barrier, $\langle E \rangle = \langle E_e \rangle \approx \frac{3}{2} kT$. The simplest approach relates the momentum, p, of an ultrarelativistic particle inside a black hole, in this case a square well of width $2R_H$, to its de Broglie wavelength, $\lambda$, where $\lambda / 2 \approx 2R_H$. Hence

$$p = \frac{h}{\lambda} \approx \frac{h}{4R_H} = \frac{h}{4\left(\frac{2GM}{c^2}\right)} = \frac{hc^2}{8GM} \quad . \tag{7.4}$$

Combining eqs. (7.3) and (7.4):

$$T \approx \frac{\langle E_e \rangle}{\frac{3}{2}k} = \frac{2E}{3k} = \frac{2pc}{3k} = \left[\frac{hc^3}{12kG}\right]\frac{1}{M} \quad . \tag{7.5}$$

The Hawking 1974 value for temperature is a factor of 2 smaller than his 1975 value. This is not critical, and the 1975 expression is

$$T = \left[\frac{\hbar c^3}{4\pi kG}\right]\frac{1}{M} = \left[2.46 \times 10^{23}\right]\left(\frac{1}{M}\right) {}^\circ K \quad , \tag{7.6}$$

with M in kg. It is remarkable that the simply derived eq. (7.5) is so close to the much more difficult derivation of eq. (7.6), differing only by the ratio $4\pi / 12$. For M ~ $10^{15}$ gm (the smallest mass that can survive to the present for Hawking), T ~ $10^{11}$ K. As we shall see, my theory permits the survival of much smaller masses such as for example M ~ $10^9$ gm with T ~ $10^{17}$ K.

**7.2 Black holes, Liouville's theorem and the ergodic hypothesis**

In the absence of collisions, the Boltzmann transport equation can yield Liouville's theorem. Liouville (1837) showed that the measure of a set of points is an invariant of the natural motion in phase space. In other words, if any portion of phase space is densely and uniformly filled with moving points representing a dynamical system in different possible states of motion, then the laws of motion are such that the density of these points remains constant in phase space. Each system is represented by a single point in phase space, and the ensemble of systems corresponds to a swarm of points in phase space.

In relating statistical mechanics to standard thermodynamics, each point represents a system of N particles, where conventionally N ~ $10^{23}$ in a phase space of 3N spatial coordinates and 3N momentum coordinates. Thus ordinarily, we might expect a need for a large number of interacting black holes. Although all the members of the ensemble should be as like our system of interest (a black hole) as permitted by physics and our knowledge, they may have any initial conditions that are allowable. The many replicas of the system of interest occupy all the admissible initial conditions and times. The volume of phase space between two energy surfaces E and E + $\Delta$E has the dimensions of (energy x time)$^{3N}$. Since the Schwarzschild black hole solution is only for one body characterized by only a few variables such as mass, angular momentum, and charge, black holes themselves seem not to be candidates for thermodynamic attributes such as temperature and entropy. Nevertheless, the black hole collective constituents such as strings or other Planck sized objects may qualify.

Since Liouville's theorem is so central to statistical mechanics, which in turn represents a foundation for thermodynamics, let us consider its applicability and limitations with respect to black hole thermodynamics. First of all Liouville's theorem applies only to a non-dissipative system in which energy (KE + PE) is conserved. To a good approximation energy is conserved for a large black hole as it hardly radiates, but this is not a good approximation for a small black hole due to Hawking radiation. The system of any size black hole and Hawking radiation conserves mass-energy. However, this is not conserved for the black hole itself.

Second and perhaps even more importantly, correlation with thermodynamics depends subtly on the ergodic hypothesis, since it is assumed that the system can indeed move from one region of phase space to another based upon the equations of motion, consistent with conservation of energy. The ergodic hypothesis implies that every state of the system can be reached directly or indirectly from every other state. This means that if the energy of the system is determined within a range $\Delta$E, the probability of finding the system in a certain state compatible with that energy is the same for each state.

Third and most importantly in not only the ergodic hypothesis, but as the basis of statistical mechanics, the time average over the evolution of the system is replaced by the average over the different states. Black hole thermodynamics in general, and black hole entropy in particular does not clearly specify what the states are. Nor is the validity of doing this self-evident in general relativity, because of the elasticity of space-time. How is it applied to the surface of a black hole where time appears to stand still from the perspective of a distant observer?

**7.3 Entropy discussion**

Black hole entropy is an ingenious creation of Jacob Bekenstein to prevent black holes from decreasing the entropy of the universe. Like the classical notion of entropy, it works exceedingly well, but may not yet be the final word on this subject. It is noteworthy that one may start with the standard extensive equation for entropy and end up with the Bekenstein entropy equation for black holes that is not extensive because of the inverse relation between temperature and mass in black holes (Rabinowitz, 2005). In reality black hole entropy is a paradox within a paradox. While solving one antinomy, it creates a new mystery. Even though everything else can disappear inside a black hole, entropy cannot disappear inside a black hole. It lurks on its exterior. In the process, a new why is created. No one knows why the entropy is associated with just the surface of a black hole, and not its entire volume. In 3-dimensional space Bekenstein (1972, 1973, 1974) found that the entropy of a black hole is

$$S_{bh} = kAc^3/4G\hbar = \frac{kc^3}{4G\hbar}4\pi R_H^2 = \frac{\pi kc^3}{G\hbar}\left[\frac{2GM}{c^2}\right]^2 = 4\pi k \frac{M^2}{\left(\frac{c\hbar}{G}\right)}, \quad (7.7)$$

$$= 4\pi k \left[\frac{M}{M_{Pl}}\right]^2 = k \ln N_s$$

where A is its surface area, M is the mass of the black hole, $M_{Pl} \equiv (c\hbar/G)^{1/2} = 2.18 \times 10^{-8}$ kg is the Planck mass, $k = 1.38 \times 10^{-23}$ J/K, and $k \ln N_s$ is the standard Boltzmann statistical mechanical entropy of a system containing $N_s$ distinct states. Eq. (7.7) has been solidly defended with theory and good thought experiments. In fact many such thought proofs have contended that entropy is associated with the area of all kinds of horizons -- not just that of black holes. However, all these arguments have a certain degree of unavoidable circularity to them since they need to be self-consistent within a given framework. Gedankend experiments are good when physical experiments are not possible, but are no substitute for real experiments .

It is not clear what distinct black hole states are being counted by $N_s$ in the expression $S_{bh} = k \ln N_s$ (Rabinowitz, 2005). A further problem is that since entropy and temperature are statistical quantities dealing with many bodies, what does it mean to speak of them with respect to a black hole viewed as a single body, which is all that the Schwarzschild solution deals with. A given black hole appears to have only 1 state that is unconditionally characterized by its mass M, angular momentum L, and charge Q. With $0 = Q = L = M$, there is no black hole, and no states and the RHS of eq. (7.7) is consistent with the LHS. With $M \neq 0$, there would appear to be only 1 state making the RHS = ln 1 = 0, which is inconsistent if not totally incompatible with the LHS which is $\propto A \neq 0$.

Black hole entropy appears to violate the Nernst heat theorem that the entropy S → 0 as the temperature T → 0. As a black hole gets larger, its temperature approaches 0, but its entropy gets arbitrarily large rather than going to 0. This appears paradoxical since the Nernst Heat Theorem is derived quite generally. In Fermi's book (1956) he says:

> ...the only circumstance under which Nerst's theorem might be in error are those for which there exist many dynamical states of lowest energy. But even in this case, the number of such states must be enormously large (of the order of ~exp N) if deviations from the theorem are to be appreciable. Although this is not theoretically impossible, it seems extremely unlikely that such systems actually exist in nature...

I think that the answer to this enigma lies neither in the semi-classical nature of black holes as we know them, nor with a quantum theory of gravity. Neither does this paradox arise because of any lack of quantum mechanics in the derivation of the Nernst heat theorem, which is deduced from quantum statistics and depends on the concept of discrete quantum states. It clearly arises from the inverse relationship between temperature and black hole mass which is quantum mechanical at its core. This gives rise to a negative heat capacity which other systems also have (Rabinowitz, 1999b, 2003). In simple terms, the Nernst heat theorem says that as temperature falls for a body with positive heat capacity, its energy decreases until at 0 K, the body is in a state with lowest possible energy and hence 0 entropy. So the Nernst heat theorem does not apply to bodies with a negative heat capacity like black holes. Even for bodies with a positive heat capacity, if the temperature goes to 0 as the density goes to 0, then the entropy need not go to 0. Interestingly, the density of a black hole goes to 0 as its mass goes to infinity:

$$\rho = \frac{M}{\left(\frac{4\pi}{3}\right)R_H^3} = \frac{M}{\left(\frac{4\pi}{3}\right)\left(\frac{2GM}{c^2}\right)^3} = \frac{3c^6}{32\pi G^3 M^2} \xrightarrow[M \to \infty]{} 0 \qquad (7.8)$$

It is noteworthy that the simplest expression for the power radiated from a black hole is in terms of its density. If we combine the Hawking (1974, 1975) power with eq. (7.8) we obtain

$$P_{SH} = \frac{\hbar c^6}{960\pi G^2 M^2} = \frac{G\rho\hbar}{90}. \qquad (7.9)$$

Similarly, if we combine the GTR power with eq. (7.8) we obtain

$$P_R = \frac{\hbar c^6}{16\pi G^2 M^2} \frac{\left\langle e^{-2\Delta\gamma} \right\rangle}{M^2} = \frac{2G\rho\hbar \left\langle e^{-2\Delta\gamma} \right\rangle}{3}. \tag{7.10}$$

Equations (7.9) and (7.10) say intuitively that the radiated power increases linearly as the density of the black hole increases. We may similarly obtain the temperature of a black hole by combining eqs. (7.6) and (7.8):

$$T = \left[ \frac{\hbar c^3}{4\pi k G} \right] \frac{1}{M} = \frac{1}{k}\sqrt{\frac{2\rho}{3\pi}}. \tag{7.11}$$

The area of a black hole in terms of its density is

$$A = 4\pi R_H^2 = \frac{3M}{\rho R_H} = \frac{3M}{\rho\left(\frac{2GM}{c^2}\right)} = \frac{3c^2}{2\rho G}. \tag{7.12}$$

Hence from eqs. (7.7) and &.(7.12) the entropy of a black hole is

$$S_{bh} = kAc^3 / 4G\hbar = \frac{kc^3}{4G\hbar} 4\pi R_H^2 = \frac{kc^3}{4G\hbar}\left(\frac{3c^2}{2\rho G}\right) = \frac{3kc^5}{8\rho G^2 \hbar}. \tag{7.13}$$

Black hole "no-hair" theorems state that time-independent, non-rotating symmetrical black holes can be completely characterized by a few variables such as mass and charge associated with long-range gauge fields. These theorems indicate that in the process of collapse, asymmetries will be radiated away essentially as gravitational radiation. Therefore, the symmetry breaking by the presence of a second body which causes the black hole to become irregular, presents a predicament in the survival lifetime of little black holes. The simple gravitational radiation fix cannot remove the perturbation and restore symmetry without the entire evaporation of the black hole (Rabinowitz, 1999b). Samir Mathur (2003) introduces black hole hair by proposing that the interior of a black hole is "not empty space with a central singularity," but that the states contributing to the entropy are bound states of branes [which] have a size that is of the same order as the horizon radius of the corresponding black hole. This brane approach has commonality with the prior work of Madacena.

### 7.3.1 *Maldacena*

Juan Maldacena's (1996) physics doctoral thesis was seminal in its exploration of black holes in string theory. He feels that "at least as far as entropy calculations is concerned" strings account well for black holes. However, he maintains some reservations with respect to other black hole properties. He views Hawking radiation as the collision of two oppositely moving open strings attached to the branes that decay into a closed string that leaves the brane. Yet he leaves the door open that strings may not lead to Hawking radiation by saying: "It will be interesting to study these more dynamical questions to understand better whether this object really represents a black hole or not."

On the question of the LIP, Maldacena feels that if his model is qualitatively correct, there should be no information loss. His model implies that the information remains on the open strings attached to the branes at the horizon. However he stipulates a qualification that, "Because of our lack of control on the strong coupling problem we cannot say anything definite about information loss." Nevertheless he feels that black holes are to quantum gravity, what the simple hydrogen atom was to quantum mechanics.

Although Maldacena's results are encouraging, we should bear in mind that to date there is no experimental evidence whatsoever to two critical features of string theory -- namely, higher dimensions, and supersymmetry. Attempts to demonstrate higher dimensions by finding a deviation from the $1/r^2$ gravitational field law at small distances have failed; and it is not possible to probe the curled up higher dimensions with attainable energies. The predictions of a spectrum of supersymmetric particles have also not been observed.

More recently Gary Horowitz and Juan Maldacena (2004) have invoked entanglement to attack the LIP. In the Hawking radiation model, one particle of a virtual particle pair near a black hole falls inside the horizon and its anti-particle goes outward, giving the appearance of coming from the black hole. They argue that because of entanglement, information is encoded in the outward moving particle, and can be largely unlocked. Their controversial stratagem requires a unique state for the annihilation of matter at the singularity. This mechanism for resolving the LIP is not part of the conventional understanding of quantum physics, and still leads to some loss of information.

**7.3.2** *Wald*

Wald (1992) was among the first to point out what he called major puzzles that remain with respect to the notion of black hole entropy.

> a) What is the mechanism by which thermal equilibrium can be achieved between a black hole and a material body? Since in the absence of Hawking radiation, a black hole cannot causally influence its exterior?

b) Underlying ordinary thermodynamics and the interpretation of entropy is the idea that "time average = phase [space] average."  How is this applied to a black hole, or what idea replaces it?

c)  Why is the entropy of a black hole so simply and directly related to its horizon area, even in the non-equilibrium state?

## 8  Other Possible Inconsistencies or Paradoxes

A paradox is a kind of inconsistency.  There are additional possible inconsistencies specifically related to black holes, and some more general ones directed at EGR itself. It is noteworthy that when Fischbach et al (1986) proposed a Fifth Fundamental Force to slightly counteract gravity at intermediate ranges, no one seems to have been troubled that it would have contradicted the equivalence principle and hence the foundation of general relativity.  This dire consequence did not have to be faced, as the fifth force faded far away.

### 8.1  The speed of light

Richard Feynman (1985) allowed that light could go much faster than the established speed limit of light.  Despite the apparent contradiction with relativity,  he seemed not to be worried about it in saying:

> ... but there is also an amplitude for light to go faster (or slower) than the conventional speed of light.  You found out that in the last lecture that light doesn't go only at the speed of light.

It has been demonstrated by conclusive experiments that photons can tunnel through a barrier with a group velocity that greatly exceeds the speed of light in vacuum (Chiao and Steinberg, 1997).  In one experiment, a photon traveled with superluminal speed through a barrier as compared with a control photon that went the same distance in vacuum.  Similar results were obtained by other groups using microwaves, and femtosecond lasers.  In one experiment an intelligible microwave version of  Mozart's 40th symphony tunneled through a barrier at almost five times the speed of light.  A consensus has not been reached as to whether or not such superluminal tunneling violates relativity.

Some argue that special relativity does *not* require the group velocity, $v_g$ , to be less than the speed of light, c, in vacuum. Sommerfeld speaks of a signal or front velocity that cannot exceed c. A *monochromatic* (single frequency) *wave* cannot carry information and has neither beginning nor end, *so its velocity can exceed c*. Phase velocities, $v_p$, can exceed c in wave guides and in the anomalous dispersion of light. When the relation $v_p v_g = c^2$ holds, and if $v_p < c$, then $v_g > c$. So it is not always clear when physical events must occur inside the light cone in obeying relativity. *The speed of gravitational tunneling may well exceed c.* The classical tunneling model of Cohn and Rabinowitz (1990) may give an insight as to how this can occur.

It has long been a theme in my deliberations that there appears to be an inconsistency between quantum mechanics and general relativity, or at least the weak equivalence principle for a number of reasons (Rabinowitz, 1990, 2001a, 2003). Examination of quantized gravitational orbits demonstrates a disparity between quantum mechanics and the weak equivalence principle because the equations of motion depend on the orbiting mass. Similarly quantum-gravitational interference effects in particular, depend on the phase which depends on the mass. Gravitational tunneling may also be at odds with the weak equivalence principle and hence general relativity. There is a discrepancy whether the gravitational tunneling occurs faster or slower between two points than a classical trajectory. There is most certainly a problem if the tunneling is superluminal as are the photon experiment described above. As I pointed out in Sec. 4.6, there even appears to be a disparity between quantum localization and black hole localization.

Because of the discrete spacetime in *loop quantum gravity* (Smolin, 2001), high energy photons travel faster than c, with low energy photons going at c. One problem with this is that the wavelength of a photon is a function of the relative velocity between the source and the observer. A bigger problem is that speeds > c fly in the face of special relativity, requiring a modification of Einstein's theory to accommodate this. In a medium with index of refraction >1, high energy photons do go faster than low energy photons, but all their speeds are ≤ c.

**8.2 The speed of gravity**

There is a well-known disparity between Newtonian gravity (NG) and EGR. Gravity propagates instantaneously in NG, i.e. the speed of gravity is infinite in NG. Never mind that this is not consistent with Newton's conviction that distant bodies must act on each other through intermediaries and that there is no action-at-a-distance. But we need to bear in mind that an infinite speed of gravity is clearly inconsistent with the conclusion of EGR that gravity propagates at the speed of light. Even the latter is not without challenge as we will see at the end of this section. Contary to many popularizations, if the Sun exploded (isotropically) one would not conclude from NG that the earth would feel it instantaneously and that we would see it 8 minutes later. As long as the center of mass of the Sun's debris didn't change, we would feel it much later than we saw it, until the debris exceeded the orbital radius of the earth. However, if the Sun suddenly moved out of the plane of the ecliptic, NG would predict that the earth's orbit would change instantly, but EGR would say that it would take 8 minutes for the orbit to start changing.

    Because NG works so well, EGR must and does reduce to NG in the weak-field limit, and resourcefully gives the same results as NG while maintaining the speed of gravity ≤ c. Even though EGR obeys the universal speed limit of c, it as if the speed of gravity were infinite because the effect of retardation almost cancels out. It would completely cancel out if there were no gravitational radiation. It is ridiculous to think that any change in a gravitational field anywhere propagates instantly throughout the universe. Even though the gravitational field of the Sun results from its past position as a retardation effect due to the finite speed of gravity, EGR predicts that the gravitational field of the sun appears to come from its present instantaneous position, neglecting gravitational radiation. Gravitational radiation results in only a small correction; is much weaker than electromagnetic radiation; and also propagates at the speed of light . From planetary orbits down to electronic orbits in atoms, orbit instability due to gravitational radiation would not be detectable down to times of the age of the Universe ~ 14 billion years. To appreciate how small this effect is, it has been observed that gravitational radiation reduces the orbital period of a binary pulsar by only 75 microseconds/year . This observation won the 1994 Nobel Prize for Joseph Taylor and Russell Hulse.

    Just as one derives the electromagnetic propagation speed of light in any medium

$$c_{any\ medium} = 1/\sqrt{(\text{pemittivity})(\text{permeability})} \qquad (8.0)$$

by deriving the wave equation from Maxwell's equations., one derives the speed of gravity = c from EGR. The speed of light slows down depending on the permittivity and permeability of the medium in which the light is propagating, and it appears that the speed of gravity or gravitational waves is reduced in going through matter. This may be similar to the Shapiro (1964) time delay prediction of radar signals reflecting from planets which supports the prediction of EGR that the time is increased for light to travel through a gravitational field. The length of the light path is increased by the presence of matter which produces spatial curvature. In a delightful little book, Clifford Will (1986) points out that the almost unknown Duane Muhleman and Paul Reichley made a similar calculation that was published in a Jet Propulsion Laboratory report two weeks before Shapiro's submission. Interestingly, although the speed of light has been measured and found to slow down in a gravitational field, the speed of gravity has not yet been measured conclusively, as that is a much harder experiment.

There are cases where gravity appears to have infinite speed for frame dragging in EGR. In 1913 while deriving general relativity, Einstein realized that it predicts space-time frame dragging for a rotating body (Einstein Papers, 1995). This is now known as the Lense-Thirring effect (1918) which is the object of serious experimental investigation. We shall not now deliberate on the potential paradox that empty space can be dragged. This and other paradoxes might be reconciled by quantum mechanics which imbues empty space (the vacuum) with an all pervading energy and mass -- but unfortunately much, much too much. However, Lindblom and Brill (1974) found in EGR that within a rotating collapsing shell of particles whose motion is only constrained by the gravitational field of the shell, frame dragging appears to be instantaneous, i.e. with an infinite speed for gravity. However, it is not clear whether this and other similar infinite speed of gravity findings in EGR may only be artifacts (Gron and Voyenli, 1999). This model is somewhat like a shell of distant stars in the universe. The speed of gravity must not exceed c in YGR.

**8.3 The speed of inflation**

Inflation is supposed to come into play within a tiny fraction of a second after the beginning of our universe, so that distant outer-boundary regions are able to communicate at ≤ c before becoming too distant to do so for the age of the universe since the big bang. Inflation theory views the early universe as expanding outward at an exponential rate (hence the name "inflation"). With this hypothesis, many long-standing cosmological problems can be laid to rest -- such as the homogeneity of the universe. It enables us to comprehend how the diametrically opposite ends of the universe that are some 24 to 30 billion light-years apart can be so much alike (implying that they somehow communicated with each other), despite the fact that light exchanged between them can only have travelled some 12 to 15 billion light-years since the beginning of the universe (Mallove, 1988). However, there is a price to laying these and other problems to rest.

The price to be paid is that the very early universe had to fly apart much faster than the speed of light -- in apparent contradiction to relativity theory that no material objects can go faster than the speed of light (even if light can, as discussed in Sec. 8.1). This would seem to imply infinite mass and infinite energy for those objects with $v > c$, according to Special Relativity. Inflation theorists simply shrug their shoulders and say that their theory does not violate relativity, since it is not the proto-stars and proto-galaxies that were moving that fast, but rather the very fabric of space itself moving the material objects apart. They argue that the expansion is not through space, but the expansion of space itself. The distance between galaxies increases, while the size of galaxies and the bodies inside them appears not to increase. Is this because the expansion energy is relatively small compared with the binding energy of these bodies? Or is it because the expansion is only acting on the large scale space and not the inner space of these bodies? Doesn't it act on the inner space of photons, since the wavelength of primordial light clearly expands? The answer to these and related questions is not clear. It's too bad that Einstein was not alive to comment on inflation himself when it was proclaimed and almost universally accepted. One can only wonder how accepting of it he would have been.

Standard inflation theory is very hard to test experimentally. Although it neatly pushes aside the Creator's hand in setting up initial conditions with extreme precision, it lacks a deep theoretical basis. Despite its acclaim and 18 years of polishing, it still has rough edges of arbitrariness. Penrose (1987) insightfully raised a crucial question. He pointed out that because the big bang model requires an extremely small initial phase space volume, the initial entropy of the universe must be exceedingly small. This presents a serious problem for inflation theory which has been so widely accepted in spite of also employing speeds greatly in excess of the speed of light.

It has not been possible to thoroughly test the concept of a scalar field that provides enormous energy for a rapid inflation of the universe, other than just measurements of cosmic microwave background density fluctuations -- which are not all in support of inflation. The conundrum of why the early universe appears to have so little entropy may only in part have a solution in that LBH would give it a larger entropy. Their strongly repulsive radiation because of their primordial proximity would cause rapidly accelerated expansion.

**8.4  The speed of Mach's principle**

Mach's principle is that the distant stars have an influence on the local properties of matter such as inertia, and Einstein was sufficiently impressed by it that he tried to incorporate it into EGR, but succeeded in only a limited sense. EGR does include Mach's principle in some special cases (Gron and Voyenli, 1999). But this comes with some problems of its own. In EGR it is hard to see how those distant stars can impose their choice of which frames are inertial and which are accelerating with sufficient alacrity to affect rapidly changing frames, if the frame information is limited to propagate at the speed of light, unless the nearby zero-point vacuum mass/energy or nearby dark matter can come to the rescue. Aside from this problem, with respect to the rotating bucket, EGR does suggest that if the bucket is at rest and the distant stars rotate around it, its surface will be parabolic as if it were rotating.

Mach's principle implies a slight anisotropy of inertial mass due to asymmetrical arrangement of masses around it. In so far as EGR incorporates Mach's principle, it also predicts a slight anisotropy of inertial mass in different directions due to an asymmetrical arrangement of masses around it such as our galaxy. In EGR the zero of potential is only at infinity, allowing for such a slight anisotropy. Hughes et al (1960) conducted very sensitive nuclear resonance experiments that did not detect any anisotropy of inertial mass within their limits of accuracy. Drever (1961) also searched for anisotropy of inertial mass using a free precession technique, but did not find one.

Such experiments are complicated by the free fall of the earth in our solar system, and the free fall of our solar system in our milky way galaxy. Since we are in a corner of our galaxy, there is quite an asymmetry of mass assembled around any test inertial mass on earth. YGR predicts no anisotropy of inertial mass essentially because the reference level of potential can be changed with no physical consequence, since in YGR only differences in potential are consequential as in NG. However, such an experiment may not decide between EGR and YGR. There is anisotropy of the surrounding matter. Nevertheless this anisotropy may be swamped by the huge effective mass of the zero-point energy, and/or the dark matter, both of which have much greater mass and mass density than ordinary matter, and they are likely more isotropic.

**8.5 Black holes get hotter as they evaporate away**

As can be seen from eq. (7.6), black hole temperature is inversely proportional to its mass. Black holes oddly get hotter (rather than cooler) with a local decrease (due to a shrinking surface area) and global increase in entropy, the more energy they lose by radiation (evaporation). This is contrary to the temperature drop that is normally associated with evaporation. However, this is only a virtual paradox that is ascribable to quantum mechanics because as the hole shrinks, the quantum wavelength must also decrease leading to a higher momentum and hence higher effective temperature. Gravitational systems tend to exhibit such a negative heat capacity. This peculiarity is not limited to black holes, and occurs in many cases in addition to Einsteinian and Newtonian gravity.

In a Joule-Thomson expansion, one usually observes a lowering of the temperature because the total energy is conserved, and in most cases the potential energy increases with the expansion leading to a lower kinetic energy and hence lower temperature.  However, at high temperature and/or high pressure, expansion can decrease the potential energy, and increase the temperature.

In electrical field emission, when resistive heating is negligible, the emission (evaporation) of an electron may lead to cooling, no energy change, or heating of the emitter depending on whether the energy level from which it is emitted is above, equal to, or below the Fermi level.  The case of heating is suggestive of black hole heating as a black hole radiates (evaporates) away.

Hawking modeled his radiation as blackbody radiation from black holes. However, it differs in two significant respects from ordinary blackbody radiation.  It differs in that an isolated emitting black hole gets hotter rather than cooler as it emits radiation.  It also differs in that an emitting black hole, according to Hawking is not limited to emitting electromagnetic radiation, but can emit anything.  Ordinary blackbody radiation is limited to electromagnetic emission.

**8.6  Average kinetic energy of emitted particles**

When resistive heating is negligible, thermionic emission leads to cooling of the emitting body (as does ordinary evaporation) because electrons are primarily emitted at energies above the Fermi energy, $E_F$. Hawking's view is generally accepted that the temperature of a black hole is inversely proportional to its mass, and solely determined by its mass. Nevertheless, it may be informative to consider the evaporative cooling process of thermionic electron emission in case it may suggest an overlooked analogue to black hole Hawking radiation despite the fact that black holes are considered to have neither a Fermi energy nor a potential barrier which emitted particles can cross over.  We will encounter at least a small surprise in deriving the mean thermal energy of the thermally emitted electrons, so it needs to be done rigorously.   The result may be relevant to the thermal energy of black hole emitted particles, even though black hole emission is not considered to be a thermionic emission process.

 The derivation will be for thermionic emission of electrons from a metal into a field-free region, but it could just as well be for any thermionically emitted particle.  In thermionic emission, the number of electrons/m$^2$-sec arriving in the x- direction at the emitting surface with kinetic energy $\frac{1}{2}mv_x^2 \geq E_F$  in the range $dv_x$ is

$$n(v_x)v_x dv_x = \left(\frac{4\pi m^2 kT}{h^3}\right) e^{E_F/kT} e^{-\frac{1}{2}mv_x^2/kT} v_x dv_x. \qquad (8.1)$$

The Richardson-Dushman equation for thermionic emission electron number density/sec is

$$N = \frac{J}{e} = \left(\frac{4\pi m k^2}{h^3}\right)(1-r)T^2 e^{-\phi/kT}, \qquad (8.2)$$

where J is the current density, r is the reflection coefficient, and $\phi$ is the work function. Richardson derived an equation of this form in 1914, except that he got $T^{1/2}$ instead of $T^2$ because he used a Maxwell-Boltzmann distribution for the free electrons inside the emitter. In 1923, Dushman re-derived this equation using the more appropriate Fermi-Dirac distribution for the free electrons in the metal.

Each of the free electrons represented by eq. (8.1) contributes an emitted (external) electron as described by eq. (8.2).

$$N = \frac{J}{e} = \int_{v_{x0}}^{\infty} (1-r) n(v_x) v_x dv_x. \qquad (8.3)$$

Let $v_{ex} \equiv$ the velocity of an electron in the x-direction after emission.

$$\tfrac{1}{2}mv_x^2 = \tfrac{1}{2}mv_{ex}^2 + E_S \Rightarrow v_{ex} dv_{ex} = v_x dv_x, \qquad (8.4)$$

where $E_S$ is the energy difference between an electron at rest inside the metal and one at rest in vacuum. The work function $\phi = E_S - E_F$. Substituting eq. (8.4) into eq. (8.1):

$$\begin{aligned} n(v_{ex})v_{ex}dv_{ex} &= \left(\frac{4\pi m^2 kT}{h^3}\right) e^{E_F/kT} e^{-(\frac{1}{2}mv_{ex}^2 + E_F + \phi)/kT} v_{ex} dv_{ex} \\ &= \left(\frac{4\pi m^2 kT}{h^3}\right) e^{-(\frac{1}{2}mv_{ex}^2 + \phi)/kT} v_{ex} dv_{ex} \end{aligned} \qquad (8.5)$$

The velocity distribution of the emitted electrons in the x-direction is $F(v_{ex})dv_{ex}$, where

$$n(v_{ex})v_{ex}dv_{ex} = F(v_{ex})dv_{ex} N. \qquad (8.6)$$

The term on the LHS of eq. (8.6) represents the number of electrons /cm²-sec arriving at the barrier which have emitted velocity $v_{ex}$. N is the number of electrons/cm²-sec which are emitted.

Dividing eq.(8.5) by eq. (8.3):

$$F(v_{ex})dv_{ex}N = \frac{\left(\frac{4\pi m^2 kT}{h^3}\right)e^{-\left(\frac{1}{2}mv_{ex}^2 + \phi\right)/kT}v_{ex}dv_{ex}}{\left(\frac{4\pi mk^2}{h^3}\right)(1-r)T^2 e^{-\phi/kT}} \quad (8.7)$$

$$= \left(\frac{mv_{ex}}{kT}\right)\left(\frac{1}{1-r}\right)e^{-\left(\frac{1}{2}mv_{ex}^2\right)/kT}dv_{ex}$$

Eq. (8.7) shows that the velocity distribution of the emitted electrons perpendicular to the surface, has a Maxwellian form. Now we can calculate an unexpected result.

Neglecting the gravitational field, the average kinetic energy of the emitted electrons is $\langle E_x \rangle = \langle \frac{1}{2}mv_{ex}^2 \rangle$.

$$\langle E_x \rangle = \frac{\int_0^\infty \frac{1}{2}mv_{ex}^2\left(\frac{mv_{ex}}{kT}\right)\left(\frac{1}{1-r}\right)e^{-\left(\frac{1}{2}mv_{ex}^2\right)/kT}dv_{ex}}{\int_0^\infty \left(\frac{mv_{ex}}{kT}\right)\left(\frac{1}{1-r}\right)e^{-\left(\frac{1}{2}mv_{ex}^2\right)/kT}dv_{ex}}$$

$$= \frac{\int_0^\infty \frac{1}{2}mv_{ex}^3 e^{-\left(\frac{1}{2}mv_{ex}^2\right)/kT}dv_{ex}}{\int_0^\infty v_{ex}e^{-\left(\frac{1}{2}mv_{ex}^2\right)/kT}dv_{ex}} = \frac{\frac{1}{2}m\left[\frac{1}{2}\left(\frac{2kT}{m}\right)^2\right]}{\left(\frac{kT}{m}\right)} = kT \quad (8.8)$$

It is noteworthy that the average kinetic energy associated with motion in the x-direction (the emission direction) is two times bigger than from a simple consideration of the equipartition of energy principle. A heuristic way to think of this result is that the barrier filters out those particles with low velocity in the emission direction and only those with high velocity in the emission direction go over the barrier. Since this result is independent of the work function $\phi$ and the Fermi energy $E_F$, the analysis should be more applicable than only for thermionic emission.

If Hawking radiation exists, there may be an overlooked analog for the thermal energy of the emitted particles. The equipartition of energy principle can be applied to the y- and z- directions, since the y- and z-components of velocity of these particles do not change as they cross the potential barrier. Thus

$$\langle E_y \rangle = \langle E_z \rangle = \tfrac{1}{2}kT . \tag{8.9}$$

Therefore $\langle E \rangle = \langle E_x \rangle + \langle E_y \rangle + \langle E_z \rangle = kT + \tfrac{1}{2}kT + \tfrac{1}{2}kT = 2kT .$  (8.10)

The energy lost by the emitter is

$$E_{lost} = |\phi| + 2kT . \tag{8.11}$$

Let us identify $|\phi|$ with the magnitude of the potential energy of a particle of mass m at the Schwarzschild radius. So if Hawking radiation exists, the mass/energy lost per emitted particle by a black hole may be given by

$$E_{lost} \approx mc^2 + \frac{GMm}{R_H} + 2kT = mc^2 + \frac{GMm}{\left(\frac{2GM}{c^2}\right)} + 2kT = \tfrac{3}{2}mc^2 + 2kT . \tag{8.12}$$

Substituting eq. (7.6) for T into eq.(8.12),

$$E_{lost} \approx \tfrac{3}{2}mc^2 + 2kT = \tfrac{3}{2}mc^2 + 2k\left[\frac{\hbar c^3}{4\pi kG}\right]\frac{1}{M} . \tag{8.13}$$

We can find an upper limit on the maximum emitted mass by simply considering the case when $E_{lost}$ equals the rest energy $Mc^2$ of the black hole and one emitted particle of rest mass m carries all this energy. In this case eq. (8.13) becomes

$$Mc^2 \approx \tfrac{3}{2}mc^2 + 2k\left[\frac{\hbar c^3}{4\pi kG}\right]\frac{1}{M} . \tag{8.14}$$

Solving eq. (8.14) for the maximum rest mass m:

$$m \approx \tfrac{2}{3}M - \frac{m_P^2}{3\pi M} , \tag{8.15}$$

where the Planck mass $m_P = \left(\frac{\hbar c}{G}\right)^{1/2} = 2.18 \times 10^{-8}$ kg. We see that for very large M, as an upper limit m may be no larger than $\tfrac{2}{3}M$. A more

-51-

rigorous calculation would include conservation of momentum and back-reaction on the metric, which were neglected by Hawking.

The smallest black hole mass for which the Hawking calculation applies is $M = m_P$. In this case

$$m \approx m_P \left( \frac{2}{3} - \frac{1}{3\pi} \right).$$  (8.16)
$$\approx 0.56 m_P.$$

**8.7 Where's the center?**

In 1691, Newton was probably the first to make a physics argument in favor of an infinite universe that can't collapse. Based upon his theory of gravitational attraction, he reasoned that since stars can only attract each other, they would eventually all collapse to a central point if the universe is finite. However, he argued that for an inifinite universe with a roughly uniform distribution of stars, there is no central point since each point would have an infinite number of stars on all sides. This is a very reasonable argument. But like many reasonable assertions, it is possible to make an even stronger argument for the opposite view.

The argument that our universe has no center because every point is moving away from every other point is a non-sequitur. In an expanding gas, all the molecules are moving away from all the other molecules. In an expanding solid, every atom moves away from every other atom. Yet the gas and the solid have a center of mass. We are accustomed to the notion that it is not meaningful to think of the universe as having a center of mass. If the universe is a three-dimensional surface of a four-dimensional sphere (finite but unbounded), then the center of mass is outside the three-space--just like the center of mass of a balloon is not in the surface of the balloon. Similarly, if the universe is hyperbolic. But, if the universe is Euclidean (flat space) the answer is not as clear. If the universe is infinite, then one might think that any point could serve as the center, since any point has an infinite number of points on all sides of it. The cosmological principle of the *equivalence of observers* and the EGR cocept that there is no preferred point such as the center of mass of the universe have their analogues in simple Newtonian models where a given point can be considered as a central origin.

For example, if the universe is uniform and has spherical symmetry, then we can imagine an infinitude of concentric spherical shells filled with stars. Even with an infinite number of spherical shells, the center of mass still remains at the center of the first sphere. For the space surrounding each shell, its mass acts as if it is concentrated at the center of the sphere. Each shell has a uniform gravitational potential for the space which it surrounds and hence exerts no force on the bodies at smaller radii. So from this point of view, whether there are a finite or infinite number of spherical shells, stars would still collapse to a point if they weren't propelled radially outward. The argument that uniform spherical shells exert no force for inner radii is general for all inverse square laws. That is why a uniformly charged spherical shell has no electric field inside the volume that it encloses, as is well known in electrostatics.

## 9 Yilmaz General Relativity (YGR)

Hüseyin Yilmaz (1958, 1982) modified Einstein's theory of general relativity (EGR) to include a gravitational stress-energy tensor that eliminates Einsteinian black holes altogether. To my knowledge it is the only general relativistic theory that does not have a black hole horizon, and is of interest for that reason. For all the major tests of EGR, the advance of the perihelion of mercury, gravitational red shift, and the bending of starlight (the least accurately measured of the three tests) which have been done at moderate fields, Yilmaz general relativity (YGR) gives essentially the same predictions as EGR, because it differs mainly for large gravitational fields. It removes so many difficulties and paradoxes that it deserves to be considered even if it is contrary to some people's sensibilities. YGR has evolved with time from his 1958 paper to his papers in the 1970's. The main question is one of consistency with nature, not one of theoretical consistency since even EGR may have some inconsistencies such as in allowing the existence of singularities, and in the interchange of space and time coordinates inside a black hole.

For now let us only compare the success of the theoretical predictions of EGR and YGR with experimental observations, without consideration of internal consistency. This is the spirit in which the Schroedinger equation has long been judged. The Schroedinger equation works remarkably better than it should in giving highly accurate predictions of the energy levels of the hydrogen atom -- the archetype problem in quantum theory. Yet it clearly neglects spin and relativity, which individually would make non-negligible contributions. One effect is the relativistic increase of the electron's mass as its velocity increases near the proton. The other effect is the interaction of the electron's intrinsic magnetic moment with the Coulomb field of the proton. The Schroedinger equation works so well precisely because it neglects **both** of these serendipitously near-canceling effects of relativity and spin. If either one were put in by itself the theory would not work nearly as well. So an isomorphism between theoretical predictions and experiments does not always relate to either how well the theory models nature (external consistency), or its internal consistency.

Yilmaz assumed that in addition to other mass/energy sources, gravitational field energy also produces curvature of space-time by adding a "gravitational stress-energy tensor" to Einstein's equations. The non-linearity of EGR allows gravity to be a source for more gravity, so gravity can only make more gravity. In YGR the direct inclusion of a gravitational stress-energy tensor can make gravity a source of less gravity because a static gravitational field has negative energy. This leads to the prohibition of Einsteinian black holes. Whereas the positive energy of gravitational radiation leads to more gravity.

Despite much effort, no one has yet solved the two-body problem in Einstein's General Relativity. Yilmaz and his colleagues (Alley et al, 1999) would say that they never will, since it is their view that EGR is a one-body theory in which one body (*e.g.* the sun) establishes a space-time curvature (field) which determines the motion of a test body (*e.g.* Mercury) that hardly perturbs the established field. The prevailing view is that Einstein, Infeld, and Hoffmann showed that EGR is a many-body theory, though it is complicated. YGR claims to be an N-body theory. Yilmaz says that if one takes the weak field limit of EGR and considers a many body problem like the perturbation effects of planetary orbits on each other, EGR doesn't work. He asserts that the weak field limit of YGR not only gives the same perturbation effects as Newtonian gravity, but also the correct advance of the perihelion of Mercury.

On the question of the precession of Mercury as well as of the other planets, Yilmaz (1992) makes a strong statement in differentiating YGR and EGR:

> Thus according to the theory here proposed, there are **basic theoretical and experimental differences from general relativity.** For example, the 529" per century **Newtonian perturbative** advance of the longitude of the perihelion of Mercury and the 1153" per century advance of that of the Earth, etc., **cannot be predicted by a 1-body theory [e.g. EGR according to Yilmaz]. The strongest reason to question general relativity is its mathematical overdetermination which leads, among other things to these difficulties. ... the new theory [YGR] does not change much the usual methods of calculation but serves to legitimize them, since the N-body solutions actually exist in the new theory [YGR].**

YGR seems to be quite compatible with Newtonian gravity (NG) as well as having desirable features of EGR. It may also have comonality with NG with respect to singularities and Newtonian black holes. Although YGR seems to agree with NG in the weak field limit, it does not go to an infinite speed of gravity, and neither does EGR. Nor should they.

In the strong field limit, YGR should not have an escape velocity ≥ c (as does NG), or YGR would have Newtonian black holes. YGR should not agree with NG in the strong field limit -- other than also allowing N-body solutions. According to Yilmaz this is **the crucial difference** between YGR and EGR. For example YGR predicts the total 575"/century advance in the perihelion of Mercury seamlessly i.e. entirely within the framework of YGR. However (at least to date) EGR only predicts the unaccounted 43"/century and one must rely on NG to supply the additional 532"/century due to planetary perturbations to obtain the total of 575"/century.

Yilmaz is reticent about applying YGR to questions concerning the entire universe. However, Mizobuchi (1985) thinks that in YGR the Hubble red-shift does not necessarily imply that the universe is expanding. Rather than signaling an expansion, it may just be a gravitational red-shift. Since this is an important point, it needs to be nailed down, particularly in view of the experimentally claimed accelerated expansion of the universe. It is important to know if YGR really favors a steady state universe, and if so, this needs to be dealt with. This is not as bad as it may seem, since there are a host of proposed modifications of EGR to account for accelerated expansion of the universe.

The Yilmaz stress-energy tensor may be able to accommodate the contribution of all electromagnetic sources, which EGR is clearly not able to do. However, Yilmaz (1977) has obtained an exact solution for a static electric charge within YGR, that cannot be done within EGR.

It is true that what were originally thought to be purely mathematical artifacts sometimes take on a very important physical reality like the magnetic vector potential in the Aharonov-Bohm effect (1959). Should we avoid or embrace this for the curvature of space-time? Or should the view that space and time are independent concepts be retained? As aptly described by Thorne, when Minkowski first proposed it, Einstein was originally reluctant to think of space-time cojoined (Thorne, 1994). In the 1908 words of Minkowski:

> The views of space and time ... are radical. Henceforth, space by itself, and time by itself, are doomed to fade into mere shadows, and only a kind of union of the two will preserve an independent reality.

Einstein was at first reluctant to accept this point of view, since for him it was just a turbid mathematical interpretation of the otherwise physical clarity of special relativity. EGR can be cast in Euclidean space with a large number of terms in the Lagrangian (Weinberg, 1972). He has shown that most of the features of the gravitational field can be derived from its symmetry properties as is the case for all other fields in quantum theory.

This tends to support the view that gravity may not be intrinsically due to space-time curvature.

For the ancient Greeks, man was the measure of all things. Special relativity says this in spades in saying that space and time are perceived differently by observers in different reference frames. In doing this, special relativity appears to have vanquished the "aether," allowing no aether-like properties for empty space. Of course properties like the permeability $\mu_o$, permittivity $\varepsilon_o$, and the impedance of empty space $\sqrt{\mu_o/\varepsilon_o}$ = 376.7 ohm are compatible with special relativity. However, General Relativity appears to have resurrected some form of the aether by imbuing space-time with properties like the ability to act on matter and light. This is certainly the case in the polarizable vacuum (aether), Euclidean representations of General Relativity in which there is no space-time curvature such as will be discussed in Sec. 10.1.

In YGR space-time is relative to the local frame of observation. It is somewhat like Susskind's black hole complementarity discussed in Sec. 4.2. For an observer at rest in a gravitational field, YGR says that space-time is curved. For an observer in free-fall, the same region is locally flat. If space-time is absolutely curved, this imbues space with ether-like properties. A relatively curved space-time removes all vestiges of an ether from the real world. This is just the opposite of Winterberg's (2002) approach to make the aether real and to give it a central role in underlying physical reality.

We have to be careful not to imbue all mathematical or conceptual constructs with physical actuality. It is not self-evident when to do it and when not to do it. The mathematical concept of space-time may or may not be the same as the individual physical concepts of space and time. Phlogiston, once thought to be a material substance required for the explanation of heat, had to bite the bullet of unreality. But one has to be careful -- since at first atoms seemed no more than a conceptual construct, and even the great Ernst Mach was reluctant to accept their reality. Action-at-a-distance has come and gone, and come again --for example in "entanglement" within quantum mechanics

## 10 Other Theories of General Relativity

### 10.1 Dicke's general relativity

Robert H. Dicke (1961) developed a polarizable vacuum (aether), Euclidean representation of General Relativity in which there is no space-time curvature. It derives gravitation as a manifestation of electromagnetism and would thus be a giant step forward toward a unified field theory if it were correct. It considers the implications of a greater aether density in the vicinity of a gravitating body. In his theory a body falls toward regions of greater gravitational field because the charged particles of which the body is composed move to the region of space where the dielectric constant of the aether is greater. This is somewhat similar to Eddington's concept of a Euclidean space-time refractory medium which gets denser and whose index of refraction increases as a gravitational source is approached.

Thus the gravitational bending of light, gravitational red shift, and the advance of Mercury's perihelion may be interpreted as related to pure electromagnetic interaction with an underlying medium of variable index of refraction -- though with a faster than light speed (Van Flandern, 1998). Although Dicke's general relativity (DGR) has an entirely different basis than that of Yilmaz, the metric that Dicke obtains is precisely the same as Yilmaz. Hence it also has no Einsteinian black holes, and makes the same predictions for the three major experimental tests of GR. For all three approaches -- EGR, YGR, and DGR -- a light wave generates twice the gravitational field per unit energy density giving twice the deflection of a light ray than expected from Newtonian gravity. As with YGR, DGR seems to have been totally neglected by the major chroniclers of GR such as Misner, Thorne, and Wheeler (1973), Wald (1984), and other texts.

**10.2 Brans-Dicke general relativity (BDGR)**

Dicke and his student Carl Brans' new theory fared better, and was considered a major challenge to EGR for a while. The Brans-Dicke general relativity (BDGR) theory was an attempt to better incorporate Mach's principle than does EGR. It was motivated to clearly comply with Mach's principle. In the Brans-Dicke (1961) GR theory, masses throughout the universe generate a scalar potential field, in addition to Einstein's curvature of spacetime, which can influence the strength of the universal gravitational constant G in space and time since G is linked to the matter-energy distribution of the universe. Whenever the scalar field is high, G is low. Dicke thought that solar oblateness might in part account for the advance of the perihelion of mercury, but this did not pan out.

Dicke led theoretical and experimental challenges to GR. At the same time that he was promoting BDGR, he helped with experiments that brought about its downfall, and bolstered EGR. It is not clear what bearing these experiments had on YGR. Dicke was assisted by Peter Bender, Carroll Alley, et al. in lunar laser ranging experiments to differentiate between EGR and BDGR in 1975. They measured the approximately 2.5 sec. round trip time for a laser pulse from the earth to an array of corner reflectors on the moon (placed there for this purpose by our astronauts) within a $10^{-9}$ sec precision.  This amounted to determining the $3.8 \times 10^5$ km  (238,000 miles) orbital distance of the moon from the earth to within 15 cm (less than 1 foot).

This extraordinary experiment involving single photon detection was instrumental in the ascendance of EGR over competing theories, though it did not discriminate between EGR and YGR.  It appears that presently the accuracy can be brought down to ~ 1 cm, and that ~ 1 mm may be possible in the not too distant future -- permitting discriminatation between EGR and YGR. The precision of this experiment permitted the investigation of a myriad of important basic physics questions (Alley, 1982).  Among them, they determined that the earth and the moon fall toward the sun with the same acceleration to ~ less than a few parts in $10^{-13}$, and that if there is a secular variation of G (as Dirac had suggested in his big numbers paper) it is ~ less than a few parts in $10^{-12}$.

## 11 Conclusion

Einstein was troubled by black holes and the paradoxes that they can spawn.  Thirteen years after he submitted Schwarzschild's 1916 black hole paper derived from EGR, Einstein (1939) denied that such objects can exist in the real world, even though they are theoretically possible in general relativity.  He may have finally accepted them, but not without much trepidation.

Though black holes were long considered to be a fiction, their existence  appears now to be firmly established in the eyes of the scientific community.  In our own galaxy and in the galaxy NGC 4258, the central dark mass seems to be a black hole.  In the case of our galaxy, recent measurements of the velocities of stars as close as 5 light-days from the dynamical center may imply a black hole of  $2.6 \times 10^6$ solar masses, as reported by Genzel (1998).   There is also evidence that stars are being ripped apart by supermassive bodies.  These are likely black holes, whose great tidal force squeezes and stretches nearby stars until they disintegrate.  Observations of x-rays coming from the center of galaxy, RX J1242-11, by two orbiting x-ray telescopes, NASA's Chandra and the European XMM-Newton satellites, were indicative that a black hole of mass of about 100 million suns was tearing apart a star in its proximity.

A star that has depleted its nuclear fuel  cannot avoid gravitational collapse.  Chandrasekar  showed that EGR predicts that black holes are the final destiny of all stars with mass $\geq$  1.4 $M_{sun}$.  That is why all white dwarf stars have mass < 1.4 $M_{sun}$.  The star will end  as a neutron star if the core has a mass greater than the Chandrashekhar limit of 1.4 $M_{sun}$ and less than about 3-5 $M_{sun}$. Instead of Einsteinian black holes, YGR has "grey holes" where the emitted light is greatly red-shifted. If neutron stars with mass much greater than 3 solar masses, or white dwarfs with mass much greater than 1.4 solar masses were detected, this would favor YGR over EGR (Alley et al, 1999).

If the horizon of Einsteinian black holes doesn't exist, there is no Hawking radiation, since it depends on the existence of a black hole horizon. This possibility impinges more on Hawking's model of black hole radiation than it does on my field emission-like model of gravitational tunneling radiation. Tunneling radiation would still be nearly the same between a very dense little grey hole and a second body. If there are black holes of all kinds, there may still be no Hawking radiation. And even if Hawking radiation exists, it is generally agreed that highly charged extreme Reissner-Nordstrom black holes do not Hawking radiate as discussed in Sec. 5.3.

So resolutions of the LIP that depend on Hawking radiation, are left in the lurch. This does not happen with GTR's resolution of the LIP. Furthermore, everything falling into a BH is kept track of by the rest of the universe. Although hard to retrieve, this bookkeeping of information is not lost. Not only is the Hawking radiation derivation questionable because it deals with arbitrarily large energies close to the horizon, but it is also in conflict with loop quantum gravity and all other theories which have a shortest length or cut-off such as at the Planck length, since he assumes vacuum modes exist with wavelengths <<< the Planck length.

It is noteworthy that there are at least two contrary dictionary definitions of the word "paradox." Among the dictionary definitions of paradox are:

   a) A statement that is seemingly contradictory or opposed to common
       sense and yet is perhaps true.
   b) a self-contradictory statement that at first seems true [but is false]. c) an argument that apparently derives self-contradictory conclusions
       by valid deduction from acceptable premises.

All of these definitions apply to black hole paradoxes including those related to Hawking radiation.

Physics is about consistent modeling of nature and reality. Sometimes inconsistencies or paradoxes arise that are virtual and reconcilable. Physics cannot afford to ignore those paradoxes that may not be resolvable. We do not yet know if all the black hole paradoxes, or the paradoxes of EGR in general are resolvable. In fact there is an anomaly known as the Sagnac effect that spans both special relativity and EGR, and appears not to have been satisfactorily resolved to this day. Sagnac (1913) claimed that his rotating interferometer experiment proved that even if linear motion is relative, rotational motion is absolute in violation of special relativity. Since rotation involves acceleration, EGR seems to be called for. However, attempts to reconcile the Sagnac effect within the context of EGR are not fully accepted. Despite the fact that the Sagnac effect has been experimentally confirmed and is fully accepted, it is not even discussed in general relativity texts such as those of Misner, Thorne, and Wheeler (1973), and Wald (1984).

At its simplest level, physics only tries to describe nature. At deeper levels with more complex models, physics tries to predict and explain. Although modeling may be attempted at any given level of reality, it may not be able to go beyond certain levels or domains of validity for intrinsic reasons. To me, physics strives to be a logical system like mathematics, with one more requirement that other logical systems don't have. That additional requirement is an isomorphism with physical reality -- a one-to-one correspondence between the elements in the system and what we call the real world (Rabinowitz, 2001b). I don't subscribe to the Copenhagen view that reality is only what we can measure. I think reality is accessible to us at a deeper level than this. Some think of reality as direct experience such as experiments. But experiments depend on theory for interpretation. This is what is meant by saying experiments are theory-laden. Physics tries in part to answer the question, "What is reality?" Nobel laureate, philosopher-poet Octavio Paz (1942) said it beautifully :

> Reality, everything we are, everything that envelops us, that sustains, and simultaneously devours and nourishes us, is richer and more changeable, more alive than all the ideas and systems that attempt to encompass it. ... Thus we do not truly know reality, but only the part of it we are able to reduce to language and concepts. What we call knowledge is knowing enough about a thing to be able to dominate and subdue it.

But nature resists being subdued by showing different sides to different observers. The complementarity of special relativity is that space and time are different for an observer in a rest frame than they are for a relatively moving observer. The two will not agree on their measurements. In general relativity, time is different for stationary observers in different gravitational fields. There is also complementarity in quantum theory where a physical situation can only be described imprecisely in terms of a pair of somewhat mutually exclusive (canonically conjugate) variables as in the Uncertainty Principle. So black hole complementarity is compatible with these traditions. However, if Hawking radiation does not exist - - there is no evidence that it does -- then the arguments for black hole complementarity may vanish with it.

## Acknowledgment

I wish to thank Frank Rahn, Laverne Rabinowitz, and John Williams for helpful comments.

## References

Aharonov, Y. and Bohm, D. *Phys. Rev.* (1959) **115**, 485.

Alcubierre, M., Gonzalez, J.A., Salgado, M. Sudarsky, D. *ArXiv* (2004) gr-qc/0406070

Alley, C.O., *Quantum Optics*, eds. P. Meystre and M.O. Scully. Plenum, NY (1982).

Alley, C.O., Leiter, D., Mizobuchi, Y., and Yilmaz, H. *ArXiv* (1999) astro-ph/9906458.

Argyres P C, Dimopoulos S, and March-Russell J. *Phys.Lett.*(1998) **B441**, 96.

Balbinot, R. *Classical & Quantum Gravity* (1986) **3**, L107 .

Balbinot, R., Fabbri, A., and Shapiro, I. *Phys. Rev. Lett.* (1999) **83**, 1494.

Bardeen, J. M., Carter, B. and Hawking, S.H. *Commun. Math. Phys.* (1973) **31**, 161 .

Bekenstein, J. D. *Nuovo Cimento Letters* (1972) **4**, 737.

Bekenstein J. D. *Phys. Rev.* (1973) **D7**, 2333.

Bekenstein, J. D. *Phys. Rev.* (1974) **D9**, 3292.

Belinski, V. A. *Phys. Lett.* (1995) **A209**,13.

Bohm, D. *Quantum Theory*, Prentice-Hall, N.J. (1958).

Brans, C. and Dicke, R.H. *Phys. Rev.* (1961)**124**, 925.

Chandrasekar, S. *Astrophysical Journal* (1931) **74**, 81.

Chandrasekar, S. *The Mathematical Theory of Black Holes*, Oxford Univ. Press, N.Y. (1983).

Chiao, R. Y. and Steinberg, A. M. in *Progress in Optics XXXVII*, ed. E. Wolf, (Elsevier, Amsterdam, 1997), p. 345.

Cohn, A. and Rabinowitz, M. *Intl. J. Theo. Phys.* (1990) **29**, 215.

Davies, P. *Journal of Physics* (1975) **A8**, 609.

Davies, P. *The New Physics*, Cambridge Univ. Press, Great Britain (1992).

De Sabbata, V. and Sivaram, C. in *Black Hole Physics*, eds. De Sabbata, V., and Z. Zhang, Kluwer Acad. Publ., Boston. (1992).

Dicke, R.H. in *Proc. of the Int'l School of Physics, Enrico Fermi Course XX: Evidence for Gravitational Theories*, ed. C. Moller (Italy 1961) p. 1.

Drever, R.W. *Phil. Mag.* (1961) **6**, 683.

Einstein, A. *Preuss. Akad. Wiss. Berlin* (1915) **47**, 831.

Einstein, A. *Annals Math.* (1939) **40**, 922.

Einstein, A. *The Collected Papers of Albert Einstein*, **5**, Princeton Univ. Press,

   NJ 1995).
Fermi, E. *Thermodynamics* (1936) Repub.Dover New York, N.Y. (1956).
Feynman, R.P. *Quantum Electrodynamics*, Princeton Univ. Press, N.J. (1985).
Finkelstein, D. *Physical Review* (1958) **110**, 965.
Finkelstein, D. *Quantum Relativity*. Springer-Verlag, Heidelberg (1996).
Fischbach, E., Sudarsky, D., Szafer, A., Talmadge, C., and Aronson, S.H. *Phys. Rev. Lett.* (1986) **56**, 3.
Frolov, A.V. *Stanford University Report* SU-ITP-04-17 (2004).
Gasperini, M. and Veneziano, G. *Physics Reports* (2003) **373**, 1.
Genzel, R. *Nature* (1998) **391**, 17.
Gibbons, G.W. and Hawking S. W. *Physical Review* (1977) **D15**, 2738 .
Gron, O. and Voyenli, K. *Found. Phys.*(1999)**29**, 1695.
Harada,T., Iguchi, H., Nakao, K. *Phys.Rev.* (2000) **D62**, 084037.
Harada,T., Iguchi, H., Nakao, K. *Prog.Theor.Phys.* 107 (2002) 449.
Harada,T., Nakao, K. *Phys.Rev.* (2004) **D70**,041501.
Hartle, J. B. and Hawking, S. W. *Physical Review* (1976) **D13**, 2188.
Hawking, S. W. *Monthly Notices Royal Astronomical Soc.* (1971) **152**, 75 .
Hawking, S. W. *Nature* (1974) **248**, 30.
Hawking, S. W. *Commun. Math. Phys.* (1975) **43**, 199.
Hawking, S. W. *Phys. Rev.* (1976) **D14**, 2460.
Hertog, T., Horowitz, G.T., Maeda, K. *Phys.Rev.Lett.* (2004) **92**, 131101.
Horowitz, G.T. and Maldacena J. *J. High Energy Phys.* (2004) **2**,8.
Hughes, V.W. , Robinson, H.G., and Beltran-Lopez, V.B. *Phys. Rev. Lett.* (1960) **4**, 342.
Jacobson, T. *Phys. Rev.* (1998) **D57**, 4890.
Joshi,P.S., Dadhich, N. and Maartens, R. *Phys. Rev.* (2002) **D65**, 101501 .
Khalatnikov, I.M. and Lifshitz, E. M. *Phys.Rev.Lett.* (1970) **24**, 76.
Laplace, P.S. *Allgemeine Geographische Ephemeriden* (1799).[Proof of his 1795-6 claim.]
Lense, J. and Thirring, H. *Phys. Z.* (1918) **19**, 156.
Lindblom, L. and Brill, D.R. *Phys. Rev.* (1974) **D10**, 3151.
Liouville, J. *Journal de Mathematiques Pures et Appliquees* (1837) **2**, 16.
Maldacena, J. M. *PhD Thesis Harvard Univ. ArXiv* (1996) hep-th/9607235.
Mallove, E. F. *Sky and Telescope* Sept. (1988) 253.
Mathur, S. Talk given at 'Quantum Theory and Symmetries', Cincinnati, September 2003. *ArXiv* (2004) hep- th/ 0401115.
Michell, J. *Phil. Trans. Royal Soc. London* (1784) **74**, 35. [Presented Nov. 27, 1783.]
Misner,C.W.,Thorne,K.S.,and Wheeler,J.A. *Gravitation* Freeman S.F. (1973).
Mizobuchi, Y. *Hadronic Journal* (1985) **8**, 193.
Page, D., *Phys.Rev.Lett.* (1993a) **71**, 1291.
Page, D., *Phys.Rev.Lett.* (1993b) **71**, 3743.
Paz, Octavio *Poetry of Solitude and Poetry of Communion*, Senneca

   Publishing House, Mexico City, Mexico (1942).

Penrose, R. *Nuovo Cimento* (1969) **1,** 252.

Penrose, R. in *General Relativity, an Einstein Centenary Survey*, ed. S.W. Hawking and W. Israel, Cambridge Univ. Press, Cambridge (1979).

Penrose R. in *Three Hundred Years of Gravitation,* ed. S.W.Hawking and W. Israel, Cambridge Univ Press, Cambridge (1987).

Preskill, J *Proc.Intl. Symp. Black Holes, Membranes, Wormholes, and Superstrings. ArXiv* (1992) hep-th/9209058.

Rabinowitz, M. *IEEE Power Engineering Review* (1998) **18**, No.8, 24.

Rabinowitz, M. *Astrophysics and Space Science* (1999a) **262**, 391.

Rabinowitz, M. *IEEE Power Engineering Review* (1990) **10**, No. 4, 27.

Rabinowitz, M. *Physics Essays* (1999b) **12**, 346.

Rabinowitz, M. *Int'l Journal of Theoretical Physics* (2001a) **40**, 875.

Rabinowitz, M. *Journal of New Energy* (2001b) **6** ,113.

Rabinowitz M. *Focus on Astrophysics Research,* Editor L.V. Ross; NovaScience Publishers, Inc. N.Y., (2003), pp. 85 - 108.

Rabinowitz M. *Progress in Dark Matter Research* Editor J. Blain ; NovaScience Publishers, Inc. N.Y., (2005), pp. 1 - 54.

Rivas, M. *Kinematical Theory of Spinning Particles: Classical and Quantum Mechanical Formalism of Elementary Particles*, Kluwer, Dordrecht (2001).

Sagnac, G. *G. R. Acad. Sci. Paris* (1913) **147**, 708.

Schwarzschild,K. *Deut.Akad.Wiss.Berlin, Kl. Math.-Phys.Tech.*(1916) **189**, 424.

Shapiro, I.I. *Physical Review Letters* (1964) **13**, 789.

Shapiro, S.L., Teukolsky, S.A. *Phys. Rev.* (1991a) **66**, 994.

Shapiro, S.L., Teukolsky, S.A. *Am. Scientist* (1991b) **79**, 330.

Singh, T.P. *Workshop on Black Holes*, Bangalore, *ArXiv* (1998)gr-qc/9805066.

Smolin, L. *Three Roads to Quantum Gravity* Basic Books (2001).

Srednicki, M. *ArXiv* (2002) hep-th/0207090.

Stephens,C.R.,'t Hooft,G.,and Whiting, B.F.,*Class.Quant.Grav.* (1994) **11**, 621.

Susskind, L. and Thoriacius, L. *Nucl. Phys.* (1992) **B382**, 123.

Susskind, L. *Phys. Rev. Lett.* (1993) **71**, 2367.

Susskind, L. and Thoriacius, L. *Phys. Rev.* (1994) **D49**, 966.

Susskind, L. and Uglum, J. *Nucl.Phys.Proc.Suppl.* (1996) **45BC**, 115.

't Hooft, G., *Class.Quant.Grav.* (1999) **16**, 3263.

't Hooft, G., *Erice lecture , ArXiv* (2000) hep-th/0003005.

't Hooft, G., *Conf. Proceedings, "Quo Vadis Quantum Mechanics"*,Phila.(2002).

Thorne, K.S. *Black Holes & Time Warps* Norton, N.Y. (1994).

Tolman, R.C. *Relativity, Thermodynamics and Cosmology* Clarendon Press, Oxford (1958)

Unruh, W. G. *Phys. Rev.* (1976) **D 14**, 870.

Unruh, W. G. and Wald R. M. *Phys. Rev.* (1984) **D 29**, 1047.

Van Flandern, T. *Physics Letters* (1998) **A 250**, 1.
Virbhadra, K. S. *ArXiv* (1996) gr-qc/9606004.
Virbhadra, K. S. and Ellis G. F. R. *Phys.Rev.* (2002) **D65**, 103004.
Wald, R. M. *Commun. Math. Phys.* (1977) **54**, 1.
Wald, R.M. *General Relativity*, Univ. Chicago Press, Chicago, (1984).
Wald, R.M. in *Black Hole Physics*, eds. De Sabbata, V., and Z. Zhang Kluwer Acad. Publ., Boston. (1992).
Weinberg, S. *Gravitation and Cosmology*, Wiley, New York, (1972).
Will, C. *Was Einstein Right? Putting General Relativity to the Test.* Basic Books, NY (1986)
Winterberg,F. *The Planck Aether Hypothesis,* Gauss Sci.Press, NV (2002).
Yilmaz, H. *Physical Review* (1958) **111**, 1417.
Yilmaz, H. *Nuovo Cim. Lett.* (1977) **19**, 617.
Yilmaz, H. *Int'l Journal of Theoretical Physics* (1982) **21**, 871.
Yilmaz, H. *Nuovo Cim.* (1992) **107**, 941.
Zel'dovich, Y. B. *JETP Letters* (1971) **14**, 180.
Zucker M. *Physics Essays* (1999) **12**, 92.